\documentclass[twocolumn]{aastex631}
\usepackage{wasysym} 						
\usepackage{verbatim}
\usepackage{soul}
\usepackage[T1]{fontenc}
\usepackage{apjfonts} 
\usepackage[figure,figure*]{hypcap}
\usepackage{rotating}
\usepackage{multirow}
\usepackage{booktabs}
\usepackage{enumitem}

\newcommand{\approptoinn}[2]{\mathrel{\vcenter{
 \offinterlineskip\halign{\hfil$##$\cr
 #1\propto\cr\noalign{\kern2pt}#1\sim\cr\noalign{\kern-2pt}}}}}
\newcommand{\appropto}{\mathpalette\approptoinn\relax}
\newcommand{\be}{\begin{equation}}
\newcommand{\ee}{\end{equation}}

\usepackage{xspace}
\newcommand\codename[1]{\textsc{#1}\xspace}
\newcommand{\parthenon}{\codename{Parthenon}\xspace}
\newcommand{\athena}{\codename{Athena++}\xspace}
\newcommand{\kathena}{\codename{K-Athena}\xspace}
\newcommand{\kokkos}{\codename{Kokkos}\xspace}
\def\athenapk/{\codename{AthenaPK}}
\def\flash/{\codename{FLASH}}

\begin{document}

\title{The Density Distribution of Compressively-Forced Supersonic Turbulence Depends on the Driving Correlation Time}
\shorttitle{Compressive Forcing}
\shortauthors{Grete, Scannapieco, Br\"uggen, \& Pan}

\author[0000-0003-3555-9886]{Philipp Grete}
\affil{University of Hamburg, Hamburger Sternwarte, Gojenbergsweg 112, 21029, Hamburg, Germany}

\author[0000-0002-3193-1196]{Evan Scannapieco}
\affil{School of Earth \& Space Exploration, 
Arizona State University, 781 Terrace Mall, Tempe, AZ 85287, USA; evan.scannapieco@asu.edu}

\author[0000-0002-3369-7735]{Marcus Br\"uggen}
\affil{University of Hamburg, Hamburger Sternwarte, Gojenbergsweg 112, 21029, Hamburg, Germany}

\author[0000-0002-0502-8593]{Liubin Pan}
\affil{School of Physics and Astronomy, Sun Yat-sen University, 2 Daxue Road, Zhuhai, Guangdong, 519082, People's Republic of China; panlb5@mail.sysu.edu.cn}

\begin{abstract}

Supersonic turbulence plays a critical role in shaping astrophysical systems, from molecular clouds to the circumgalactic medium. Key properties of this turbulence include the Mach number, driving scale, and nature of the driving mechanism, which can be solenoidal (divergence-free), compressive (curl-free), or a mix of the two. A less studied property is the correlation time of the driving accelerations, $\tau_{\rm a}.$ While this timescale has a minimal impact on solenoidally-driven turbulence, we show that it has a strong impact on compressively-driven turbulence. Using high-resolution simulations with tracer particles, we analyze the evolution of density fluctuations, focusing on the PDF of the logarithmic density, $s$, and its rate of change, $\frac{ds}{dt},$ and the conditional statistics of $\frac{ds}{dt}$ and $\frac{d^2s}{dt^2}$. When the driving correlation time is comparable to the eddy turnover time, $\tau_{\rm a} \approx \tau_{\rm e},$ compressive driving leads to the formation of large, low-density voids in which the variance of $\frac{ds}{dt}$ is large. These are directly linked to sustained accelerated expansions, which results in a strong correlation between density and the divergence of the driving acceleration field. In contrast, when $\tau_{\rm a} \approx 0.1 \, \tau_{\rm e}$, compressive driving does not produce such voids, resulting in a narrower, less skewed distribution. We show using analytical estimates that $\tau_{\rm a}$ is may be significantly less than $\tau_{\rm e}$ in supernova-driven turbulence, highlighting the need to better understand the role of the driving correlation time in shaping the density structure of turbulent astrophysical systems.

\end{abstract}


\section{Introduction}

Supersonic turbulence is ubiquitous in astrophysics, playing a key role in protoplanetary disks, active galactic nuclei, star-forming molecular clouds, and the interstellar and circumgalactic media \cite[e.g.,][]{Brandenburg95, Schekochihin09, Moesta15, Walch15, Kim17, Buie20, Rosotti23}. In such systems, the properties of the flow depend on global conditions such as the magnetic field strength \cite[e.g.,][]{Kim03, Li04, Li15, Xu19, Seifried20, Pattle22}, the virial parameter \cite[e.g.,][]{Zweibel95, Krumholz06, Ballesteros-Paredes07, Dib07, Tasker09, Hopkins13b}, and the effective equation of state \cite[e.g.,][]{Jappsen05, Hennebelle09, Federrath15, Gray15, Zhuravleva18}. They are also dependent on properties of the turbulence itself, including the Mach number and nature of the driving mechanism \cite[e.g.,][]{Padoan97, Ostriker2001, Federrath08, Burkhart09, Pan19}.

Together, these factors determine the overall statistics of the medium, including one-point statistics, such as the density probability distribution function (PDF), and the two-point measures such as velocity structure function and power spectrum.  The statistics in turn determine the observed distribution of column densities \cite[e.g.,][]{Burkhart12, Kainulainen13}, dust extinctions \cite[e.g.,][]{Kainulainen11}, and ion abundances \citep[e.g.,][]{Buie18, Koplitz23}, as well as the observed kinematics \cite[e.g.,][]{Larson81, Brunt04, Kritsuk07, ForsterSchreiber09, Federrath13, Koplitz23}. 

The density PDF is particularly important in the case of molecular clouds, where it sets the collapse rate of dense cores, the stellar initial mass function, and the overall star-formation rate \citep[e.g.,][]{Krumholz05, Hennebelle11, Padoan11}. In numerical simulations, many of the overall characteristics of these clouds can be reproduced by studying the fundamental case of hydrodynamic isothermal turbulence \citep[e.g.,][]{Larson81, Federrath10}. 

In such simulations, turbulence is driven by accelerations on large scales, modeled  either as a static pattern \citep[e.g.,][]{MacLow98, Stone98} or, more commonly, as an Ornstein-Uhlenbeck (OU) process with finite autocorrelation timescale, $\tau_{\rm a}$, \citep[e.g.,][]{Eswaran88, Schmidt09}. The strength of the driving sets the Mach number, as large-scale motions cascade toward small scales, forming shocks and complex density structures through nonlinear processes.

In these simulations, the volume-weighted PDF of the logarithmic density can be approximated by a Gaussian,
\be
P_{\rm V}(s) \approx \frac{1}{\sqrt{2 \pi \sigma_{s,V}^2}} {\rm exp} \left[ - \frac{(s - s_{0,V})^2} {2 \sigma_{s,V}^{2}} \right],
\label{eq:PDF}
\ee
where $s \equiv \ln(\rho/\rho_0)$, $\rho_0$ is the mean density, and the mean value of $s,$ is related to the variance, $\sigma_{s,V}^2$, as $s_{0,V} = - \sigma_{s,V}^2/2$ by mass conservation \citep{VazquezSemadeni94,Padoan97,Federrath10,Padoan11}. A similar Gaussian fit, i.e., replacing subscripts $_V$ with subscripts $_M$ in eq.~(\ref{eq:PDF}), provides a good approximation to the mass-weighted PDF, $P_{\rm M},$ and in the case in which $P_{\rm V}$ is exactly Gaussian, $P_{\rm M}$ must also be Gaussian, with $\sigma^2_{s,M} = \sigma^2_{s, V}$ and $s_{0,M}=\sigma_{s,M}^2/2.$

Previous studies have shown that $\sigma_{s, V}^2 \approx \ln(1 + b^2 M_{\rm V}^2)$, where $M^2_{\rm V} = \sigma^2_v/c^2_s$ is the volume-weighted ratio of the mean velocity dispersion $\sigma^2_v$ to the sound speed $c^2_s.$  Here $b$ is a fit parameter that depends on the nature of the turbulent forcing, which is made up of two main types of motions: solenoidal and compressive.

Solenoidal (divergence-free) motions arise from processes that involve rotational forces. In the ISM this can operate in regions that are subject to a significant differential rotation\citep[e.g.,][]{Kim07,Sur16,Federrath16} as well as the magnetorotional instability (MRI) \citep{Piontek2007,Tamburro09}. They can also arise in the presence of magnetic fields, through the action of a non-vanishing Lorentz force \citep[e.g.,][]{Kahn12}, or through the conversion of compressive motions into vortical energy \citep{Brandenburg25}.

When accelerations are purely solenoidal, simulations find that $b\approx$ $1/3$ \citep{Padoan97, Ostriker2001, Price11, MacLow05, Kowal07, Glover07, Lemaster08}, and $P_{\rm V}(s)$ is very close to Gaussian for low and moderately supersonic Mach numbers, but slightly skewed toward low densities at high Mach numbers \citep{Kritsuk07, Burkhart09, Pan19}.

Compressive (curl-free) motions, on the other hand, can be driven by any process that can be described as the gradient of a potential function, such as  gravitational collapse \citep[e.g.,][] {VazquezSemadeni98,Klessen10,Elmegreen10,Robertson12}, or compression and cooling \cite[e.g.,][]{Dobbs08}.  They can also arise from supernovae and expanding radiation fronts from high-mass stars \citep[e.g.,][]{McKee89,Goldbaum11,Peters11}, which drive shocks radially outwards from small regions.

For purely compressive forcing, the variance is larger, with $b\approx 1$ providing a better fit. In this case, eq.\ (\ref{eq:PDF}) is less accurate, as $P_{\rm V}$ is significantly skewed toward low densities even at moderate Mach numbers \citep{Federrath08, Schmidt09, Konstandin12, Hopkins13a, Federrath13, Squire17}.

These one-point statistics of density fluctuations provide a picture of the steady-state distribution, but they do not capture the evolution of individual parcels of gas, which continuously change in density. An exploration of Lagrangian evolution of these parcels provides further insights into the density probability distribution.

In \cite{Paper1} we used tracer particles to track $s$ and its rate of change along Lagrangian trajectories in Eulerian simulations of supersonic turbulence with solenoidal driving. This allowed us to determine that the temporal correlation functions of $s$ and $\frac{ds}{dt}$ decay exponentially on a  timescale of $\approx 1/6$ the eddy turnover time, $\tau_{\rm e}$. It also allowed us to measure the conditional averages of $\frac{d^2 s}{d t^2}$ and $\left(\frac{ds}{dt}\right)^2,$ as a function of $s,$ which determine the shape of the density PDF \citep{Nordlund99,Pan19} and provide insights into its origin. For solenoidal driving, $\left<\frac{d^2 s}{d t^2}|s \right>/\left< \left(\frac{ds}{dt}\right)^2 | s \right>$ is nearly proportional to $-s$, with a downturn at high $s$ values due to shocks decelerating to subsonic speeds in high-density regions. This trend explains the nearly Gaussian of $P_M(s)$, with the downturn leading to the skewness at high Mach numbers.

In this work, we apply similar methods to study the effects of compressive versus solenoidal driving on the distributions of density fluctuations. In particular, 
an intriguing question to ask is whether turbulence driven with $\tau_{\rm a}$ smaller than the Lagrangian correlation time of $s$, which is about $\frac{1}{6}\tau_{\rm e}$, may exhibit significant differences from turbulence driven with $\tau_{\rm a} \approx \tau_{\rm e}$ as commonly adopted in supersonic turbulence studies. 
Furthermore, while models with short correlation times have been studied in the solenoidal case \cite[e.g.,][]{Lemaster08}, this range has been largely unexplored for compressive driving.

By carrying out compressively-driven simulations with $\tau_{\rm a} \ll \tau_{\rm e}$, here we show that the density distribution in this case is not purely a function of Mach number. Instead, $\sigma_{s,M}^2$ decreases strongly for short correlation times, an effect we study in detail using the expanded set of diagnostics allowed by the presence of tracer particles.

The structure of this work is as follows: In \S\ref{sec:sims}, we describe our numerical approach and the parameters spanned by our simulation suite. Our results are presented in \S\ref{sec:results}, including the PDF of $s$ and $\frac{ds}{dt}$ as a function of the driving mechanism and correlation time, as well as their connection with conditional averages of $\frac{ds}{dt},$ $\left(\frac{ds}{dt}\right)^2,$ and $\frac{d^2 s}{dt^2}$ as a function of $s,$ and the overall properties of the flow.
The results are discussed in the context of the driving correlation time in a supernova driven ISM in \S\ref{sec:discussion} and conclusions are given in \S\ref{sec:conclusion}.

\section{Simulations}
\label{sec:sims}

\subsection{Methods}

Our study follows the approach described in \cite{Paper1}, to produce a suite of simulations of supersonic, isothermal turbulence. Each simulation was carried out in a periodic box of size $L_{\rm box},$ over which we solved the hydrodynamic equations in the presence of a continuous stochastic driving force. The continuity and momentum equations in this case are
\begin{equation}
\frac{\partial \rho}{\partial t} + \frac{\partial \rho v_i}{\partial x_i} = 0,
\label{eq:continuity}
\end{equation} 
and 
\begin{equation}
\frac{\partial v_i}{\partial t} + v_j \frac{\partial v_i}{\partial x_j} = - \frac{1}{\rho} \frac{\partial p}{\partial x_i} + \frac{1}{\rho} 
\frac{\partial \sigma_{ij}}{\partial x_j} + a_{i} (\mathbf{x}, t),
\label{eq:velocity}
\end{equation} 
where $p(\mathbf{x}, t)$ is the pressure, $\sigma_{ij}$ is the viscous stress tensor, and $\mathbf{a}(\mathbf{x}, t)$ is the driving force. For an ideal gas, the shear viscosity is $\sigma_{ij} = \rho \nu ( \partial_i v_j + \partial_j v_i - \frac{2}{3} \partial_k v_k \delta_{ij}) $ where $\nu$ is the kinematic viscosity\footnote{
    In the present work, we only include shear viscosity as bulk viscosity is typically ignored in
    many astrophysical applications.
}. 

Unlike \cite{Paper1}, our current simulations use the \athenapk/ code\footnote{
\athenapk/ is available and maintained at \url{https://github.com/parthenon-hpc-lab/athenapk} and commit \texttt{80942e8
 } was used for the simulations.
},which implements finite volume hydrodynamic and magnetohydrodynamics algorithms on the \parthenon framework \citep{parthenon}, which is a performance-portable AMR framework based on \athena \citep{Stone20}, \kathena \citep{kathena}, and \kokkos \citep{Edwards14, Trott21}. This code offers exceptional speed and scalability, and it supports efficient AMR simulations on a range of GPUs.

For all our simulations, we employ a second-order finite volume scheme with a predictor-corrector Van Leer integrator, Harten-Lax-van Leer with Contact (HLLC) Riemann solver, and piecewise parabolic reconstruction in primitive variables. The simulations are approximately isothermal, using an ideal equation of state with an adiabatic index of $\gamma = 1.0001$. We calculate the viscous fluxes at cell faces using a second-order finite difference stencil and we integrate them in an unsplit fashion along with the Riemann fluxes. We also apply first-order flux correction in cells in which higher-order updates result in negative densities or pressures, recalculating fluxes using piecewise constant reconstruction and a Local Lax-Friedrichs (LLF) Riemann solver.

To drive turbulence, we employ a mechanical, stochastic forcing mechanism governed by an Ornstein-Uhlenbeck process \citep{Schmidt09,Grete18}. In Fourier space, this can be summarized as 
\be
\hat a_i (\mathbf{k},t + \Delta t) = c_\mathrm{drift} \hat a_i (\mathbf{k},t) + \sqrt{1-c_\mathrm{drift}^2} P_{\rm a}(k) \mathcal{P}_{ij} \mathcal{N}_j.
\label{eq:driving}
\ee
Here, $c_\mathrm{drift}=e^{-\Delta t/\tau_\mathrm{a}}$ is the drift coefficient and $\sqrt{1-c_\mathrm{drift}^2}$ the diffusion coefficient, i.e., $\tau_\mathrm{a}$ sets the correlation time of the driving, which we vary as described in detail below. $P_{\rm a}(k)$ sets the shape of the acceleration field, which we take to peak at $k_p$ according to the profile given by $P_\mathrm{a}(k) = \tilde k^2 (2-\tilde k^2) \Theta(\tilde k^2-2)$ where $\Theta$ is the Heaviside step function and $\tilde k = k/k_p.$  Finally, $\mathcal{N}_j$ are complex random numbers with modulus $<1$ and zero mean, and the 
\be
\mathcal{P}_{ij} = \left[\zeta\delta_{ij}+(1-2\zeta)\frac{k_i k_j}{|k|^2} \right]
\label{eq:driving2}
\ee
projection tensor sets the fraction of the driving power in solenoidal versus compressive modes using a Helmholtz decomposition. Here, the parameter $\zeta\in[0,1]$ determines the solenoidal fraction: $\zeta=0$ yields purely compressive driving, while $\zeta=1$ yields purely solenoidal driving.

Numerically, at each time step, we generate new, random acceleration fluctuations in spectral space whose power is proportional to the desired spectrum $P_{\rm a} (k)$. The components of these fluctuations are then projected parallel and perpendicular to the local wavevector following the desired split defined by the solenoidal weight $\zeta$. Afterward, we update the spectral acceleration field by adding the new fluctuations to the existing field weighted by the correlation time $\tau_{\rm a}$. Finally, we transform the updated field into real space, where it is normalized to ensure that no net momentum is added to the simulation and to match the target rms value, which eventually controls the overall strength of the forcing.

To track the Lagrangian evolution of mass fluctuations, we adopt a method similar to \cite{Paper1}. Each simulation includes particles initialized as a uniform lattice and evolved passively using a two-stage Runge-Kutta scheme. As in \cite{Paper1}, we employed a cloud-in-cell (CIC) mapping to linearly interpolate on a region of one cell size around each particle. We avoid higher-order mappings as they introduce unphysical tails in the $\frac{ds}{dt}$ PDF due to interpolation inaccuracies near shocks \citep{Paper1}.

For each particle, we calculated $s_n$, the value of $s$ at $t_n$ (the current time step), $s_{n-1}$ [the value of $s$ at the previous time step $t_{n-1}$), and $\frac{ds}{dt}_{n-0.5} = (s_n - s_{n-1}) / (t_n - t_{n-1})$ (the rate of change of $s$ at $(t_{n} + t_{n-1})/2$]. We also retained the values of $s$ and $\frac{ds}{dt}$ over previous times, allowing us to compute conditional averages as discussed below.

\subsection{Parameter Space}
\label{sec:parameters}

\begin{table*}[t]\centering\begin{tabular}{lcccccccccccccc}\toprule
& \multicolumn{4}{c}{Input parameters} & \multicolumn{10}{c}{Simulation properties} \\
\cmidrule(lr{.75em}){2-5} \cmidrule(lr{.75em}){6-15}
Name & $\zeta$ & $a$ & $\tau_\mathrm{a}$ & $\nu~[10^{-4}]$ & $M_\mathrm{V}$ & $M_\mathrm{M}$ & $\tau_\mathrm{a} / \tau_\mathrm{e}$ & $\left < |\nabla \times \mathbf{u}|^2 \right >_V$ & $\left < |\nabla \cdot \mathbf{u}|^2 \right >_V$ & $\nu_{\rm eff}~[10^{-4}]$ & $\eta/\Delta_x$ & $\lambda/\Delta_x$ & $\mathrm{Re}_{\lambda}$ & $\mathrm{Re}$ \\ \midrule
$\mathtt{1.0-L\nu}$ & 1.0 & 143 & 0.046 & 5.5 & 7.3 & 6.8 & 0.63 & 76,000 & 128,000 & 7.2 & 1.00 & 61 & 620 & 4,200\\
$\mathtt{1.0-S\nu}$ & 1.0 & 309 & 0.005 & 5.5 & 7.0 & 6.7 & 0.06 & 72,000 & 128,000 & 7.2 & 1.01 & 60 & 590 & 3,900\\
$\mathtt{0.3-L\nu}$ & 0.3 & 214 & 0.046 & 5.5 & 7.5 & 6.6 & 0.61 & 82,000 & 172,000 & 7.2 & 1.03 & 61 & 570 & 4,200\\
$\mathtt{0.3-S\nu}$ & 0.3 & 381 & 0.005 & 5.5 & 7.3 & 7.0 & 0.06 & 90,000 & 180,000 & 7.6 & 0.98 & 56 & 550 & 3,800\\
$\mathtt{0.0-L\nu}$ & 0.0 & 238 & 0.046 & 5.5 & 7.1 & 5.5 & 0.51 & 74,000 & 215,000 & 7.2 & 1.11 & 60 & 590 & 3,800\\
$\mathtt{0.0-S\nu}$ & 0.0 & 381 & 0.005 & 5.5 & 7.4 & 7.0 & 0.06 & 93,000 & 205,000 & 7.5 & 0.96 & 56 & 520 & 3,900\\[0.5em]
$\mathtt{1.0-L}$ & 1.0 & 143 & 0.046 & 0.0 & 7.1 & 6.8 & 0.63 & 327,000 & 178,000 & 2.3 & 0.42 & 28 & 910 & 12,500\\
$\mathtt{1.0-S}$ & 1.0 & 309 & 0.005 & 0.0 & 6.9 & 6.6 & 0.06 & 308,000 & 183,000 & 2.3 & 0.43 & 29 & 860 & 12,000\\
$\mathtt{0.3-L}$ & 0.3 & 214 & 0.046 & 0.0 & 7.6 & 6.6 & 0.60 & 305,000 & 226,000 & 2.4 & 0.45 & 32 & 910 & 12,900\\
$\mathtt{0.3-S}$ & 0.3 & 381 & 0.005 & 0.0 & 7.4 & 7.2 & 0.07 & 337,000 & 234,000 & 2.6 & 0.43 & 30 & 850 & 11,400\\
$\mathtt{0.0-L}$ & 0.0 & 238 & 0.046 & 0.0 & 7.1 & 5.9 & 0.54 & 254,000 & 246,000 & 2.3 & 0.46 & 32 & 1,030 & 12,000\\
$\mathtt{0.0-S}$ & 0.0 & 381 & 0.005 & 0.0 & 7.2 & 7.1 & 0.07 & 334,000 & 243,000 & 2.7 & 0.43 & 29 & 750 & 10,500\\[0.5em]
$\mathtt{512-1.0-L}$ & 1.0 & 143 & 0.046 & 0.0 & 7.1 & 6.6 & 0.61 & 152,000 & 88,000 & 4.9 & 0.38 & 21 & 610 & 5,600\\
$\mathtt{512-1.0-S}$ & 1.0 & 309 & 0.005 & 0.0 & 6.8 & 6.5 & 0.06 & 135,000 & 87,000 & 4.9 & 0.39 & 21 & 600 & 5,400\\
$\mathtt{512-0.0-L}$ & 0.0 & 238 & 0.046 & 0.0 & 7.3 & 6.1 & 0.56 & 115,000 & 132,000 & 5.1 & 0.42 & 25 & 650 & 5,700\\
$\mathtt{512-0.0-S}$ & 0.0 & 381 & 0.005 & 0.0 & 7.5 & 7.4 & 0.07 & 147,000 & 130,000 & 6.0 & 0.40 & 22 & 540 & 4,800\\[0.5em]
\bottomrule
\end{tabular}
\caption{Parameters of our simulations in the stationary regime.
  Columns show the run name, solenoidal fraction ($\zeta$),
  rms acceleration ($a$), correlation time of the forcing ($\tau_\mathrm{a}$),
  explicit viscosity ($\nu$), volume-weighted and mass-weighted Mach numbers ($M_{\rm V}$ and $M_{\rm M}$),
  the ratio of forcing correlation time to the eddy turnover time ($\tau_{\rm a}$/$\tau_{\rm e}$),
the volume-weighted vortical and dilatational power
    ($\left < |\nabla \times \mathbf{u}|^2 \right >_V$ and $\left < |\nabla \cdot \mathbf{u}|^2 \right >_V$),
  the effective viscosity ($\nu_{\rm eff}$),
  the effective Kolmogorov scale ($\eta$), the Taylor microscale ($\lambda$),
  and the Taylor and effective integral scale Reynolds numbers ($\mathrm{Re}_{\lambda}$ and $\mathrm{Re}$).
  The standard deviation of all simulation properties in each simulation is below $10\%$
  except for $\nu_{\rm eff}$ (and derived properties) with a maximum of $30\%$.
  For all dimensional quantities, the unit of length is the box size and the unit of time is the box sound crossing time.
  All simulations were carried out on a fixed grid of $1024^3$ cells, apart from the simulations labeled 512, which were carried out on a $512^3$ grid. The detailed definitions of the quantities are given in Sec.~\ref{sec:parameters}.}
\label{tab:runs}
\end{table*}
 
We carried out our simulations on a fixed grid with $1024^3$ cells and $512^3$ tracer particles (versus $512^3$ cells with $128^3$ tracer particles in \cite{Paper1}). The $512^3$ runs were Implicit Large Eddy Simulations (ILES), and we carried out two sets of $1024^3$ runs: one set of ILES runs and a second set with an explicit viscosity of $5.5 \times 10^{-4}$ in units of the box size and sound speed. This value was chosen to reproduce the power spectra from the $512^3$ ILES runs.

Table~\ref{tab:runs} lists the key parameters of our twelve $1024^3$ and four $512^3$ simulations, including mass-weighted and volume-weighted Mach numbers ($M_{\rm M}$ and $M_{\rm V}$). Since we wish to focus on compressive effects, we chose a fairly high Mach number in the range $\approx 6-7$ in all cases. The inferred values in Table~\ref{tab:runs} are the averages within the stationary regime, which we have taken to be the period between three and seven eddy turnover times.

At the fiducial resolution of $1024^3$, we have six simulations with viscosity and six ILES simulations. For each set, we vary the solenoidal fraction, $\zeta = 0.0, 0.3, 1.0$, and explore two driving correlation times: $\tau_{\rm a} \approx 0.6 \tau_{\rm e}$ (or L for long) and $\tau_{\rm a} \approx 0.06 \tau_{\rm e}$ (or S for short).
$\tau_\mathrm{e}$ is the eddy turnover time defined as $\tau_{\rm e} \equiv 0.5 L_{\rm box} M_{\rm M}^{-1} c_s^{-1}$
matching the definition used in \cite{Paper1}, which is $\approx30\%$ longer than another typical definition,
$\tau'_{\rm e} \equiv L_{\rm i} M_{\rm V}^{-1} c_s^{-1}$, based on the integral scale
$L_i = \int E(k)/k \, \mathrm{d}k / \int E(k) \, \mathrm{d}k (= 0.40)$ and volume-weighted Mach number.
For the S runs, these correlation times are significantly shorter than have previously been studied in cases with $\zeta < 1$.

\begin{figure*}[t]
\begin{center}
\includegraphics[width=\textwidth]{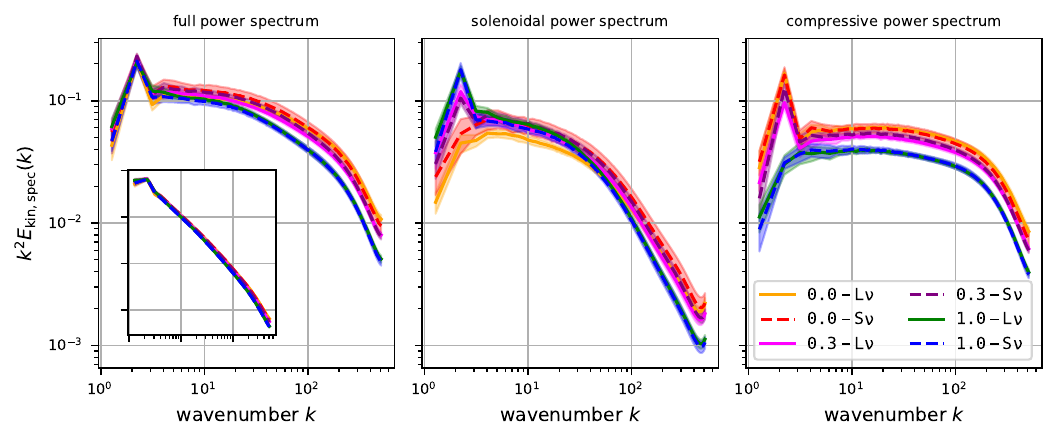}
\end{center}
\vspace{-0.1in}
\caption{Compensated specific kinetic energy spectra from simulations with explicit viscosity, showing total (left), solenoidal (center), and compressive components (right) versus normalized wavenumber. Each panel displays results from runs 0.0-L$\nu$ (orange solid), 0.0-s$\nu$ (red dashed), 0.3-L$\nu$ (magenta solid), 0.3-s$\nu$ (purple dashed), 1.0-L$\nu$ (green solid), and 1.0-s$\nu$ (blue dashed). Lines represent mean spectra over the stationary regime, and shaded regions indicate standard deviations. The left panel inset shows uncompensated spectra ($E_\mathrm{kin,spec}(k)$ vs. $k$) with the same $x$ axis range as in the compensated spectra and with the $y$-axis range from $10^{-8}$ to $10^{-1}$.}
\label{fig:Ek}
\end{figure*}

Following \cite{Pan10}, we computed the effective viscosity from the equation for kinetic energy per unit volume:
\begin{eqnarray}
\frac{\partial}{\partial t} \left\langle \frac{1}{2} \rho v^2 \right\rangle_V
& =& \left\langle p \frac{\partial v_i}{\partial x_i} \right\rangle_V + \left\langle \rho a_i v_i \right\rangle_V \nonumber \\
& & - \frac{1}{2} \left\langle \rho \nu \left( \frac{\partial v_i}{\partial x_j} + \frac{\partial v_j}{\partial x_i} - \frac{2}{3} \frac{\partial v_k}{\partial x_k} \delta_{ij} \right)^2 \right\rangle_V,
\label{eq:keperunitmass}
\end{eqnarray}
where we use the fact that the ensemble is equal to the volume average for statistically homogeneous flows. \cite{Pan19} showed that for steady-state barotropic turbulent flows the average $pdV$ work rate is zero. Therefore, we can compute the total effective viscosity by equating the viscous dissipation with the energy input rate,\footnote{
We compared the instantaneous energy input rate based on the acceleration to the actual dissipation, i.e., the
    rate of change in internal energy $\dot E_e = \Delta e / \Delta t$ with $e$ being the internal energy density
    (made possible by the use of an ideal equation of state with $\gamma = 1.0001$).
  In all simulations, both values agree to at least two significant digits.
}
$\dot E_f = \left\langle \rho a_i v_i \right\rangle_V$. The results of this analysis are shown in Table~\ref{tab:runs},
including the corresponding effective Kolmogorov scale, $\eta = \left ( \nu_{\rm eff}^3 / \dot E_e \right )^{1/4}$,
and effective integral scale Reynolds numbers,\footnote{
Note that we use the rms velocity (rather than the mean of the velocity fluctuations, which differs by
$1/\sqrt{3}$ in isotropic turbulence) as commonly done in the astrophysical literature.
} $\mathrm{Re} = M_{\rm V} c_s L_i / \nu_{\rm eff}$. Finally, we also calculate the Taylor microscale,
$\lambda = \sqrt{5 \left < |\mathbf{u}|^2 \right >_V/ \left < |\nabla \times \mathbf{u}|^2 \right >_V}$,
and associated Reynolds number, $\mathrm{Re}_\lambda = M_{\rm V} c_s \lambda / \nu_{\rm eff}$.

\section{Results}
\label{sec:results}
\subsection{Velocity Power Spectra}

Fig.~\ref{fig:Ek} shows the energy spectra averaged in time over the stationary regime. Although the uncompensated full spectra (inset, first panel) show minimal differences between runs, the compensated spectra reveal subtle variations at intermediate and small scales. These differences are related to individual power budgets, as shown in the center panel and right panels of Fig.~\ref{fig:Ek}. Here we see that the impact of driving is most pronounced on the largest scales, with the power on the driving scales directly connected to the power in the driving modes, as expected.

In the solenoidally-driven case, the energy budget is dominated by large-scale solenoidal motions. The effect of viscosity, both physical and numerical, causes the solenoidal spectra to decay around $k\approx40$ for all cases, which is similar to the energy spectrum in simulations of incompressible turbulence, where the decay starts at $k\sim k_{max}/10$ with $k_{max}$ the maximum wave number in the simulation \citep[e.g.,][]{Kaneda06}. 

In contrast, the compressive spectra are much more extended, with a sharp turnover at around $k \approx 128$. This extended $k^{-2}$ compressive spectrum likely results from shocks, and the cut-off wavelengths correspond to the shock width of several cell sizes. The more extended power law in the compressive spectrum leads to an increase in compressive-to-solenoidal ratio towards the largest wavenumbers, consistent with previous findings using different codes \citep[e.g.,][]{Federrath10, Kritsuk10}, suggesting a physical rather than numerical origin. Moreover, it may also be related to the absence of bulk viscosity in our simulations.

\subsection{Spatial Distributions}

\begin{figure*}[t]
\begin{center}
\qquad \includegraphics[width=0.96\textwidth]{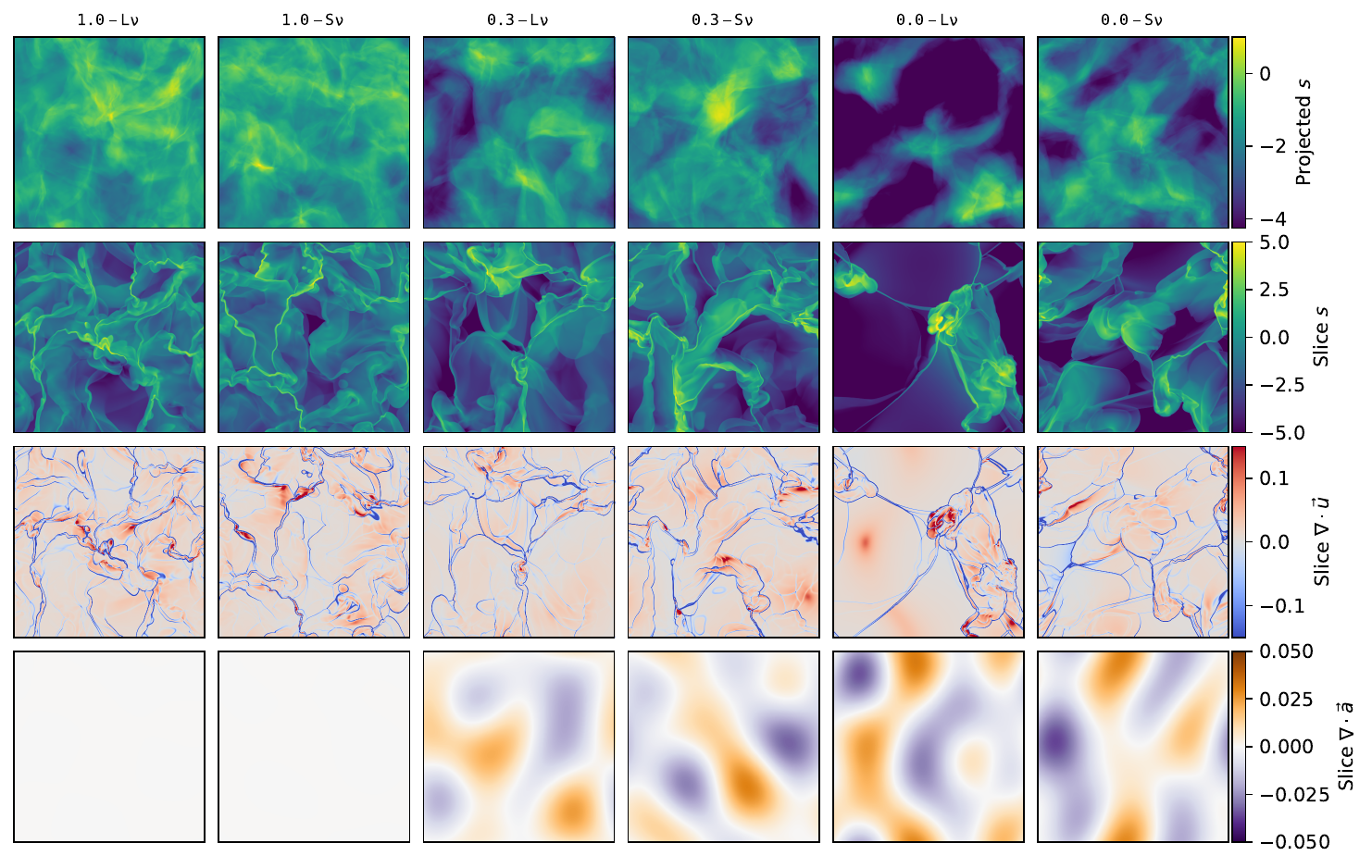}
\end{center}
\vspace{-0.2in}
\caption{Representative results from our turbulence simulations. From left to right, columns show results from runs with purely solenoidal, mixed driving, and purely compressive driving. From top to bottom, rows show the logarithm of the projected density through the box, and slices of $s$, $\nabla \cdot {\mathbf u}$, and $\nabla \cdot {\mathbf a}$. For consistency, each column shows results from the snapshot with the highest Mach number from the corresponding simulations, and in all cases, the slices are taken through the plane with the highest density. Note that the slices in the lowest row show only the divergence of the driven accelerations, not the total acceleration field.}
\label{fig:slice}
\end{figure*}

Fig.\ \ref{fig:slice} gives a visual representation of the results of our six $1024^3$ viscous simulations. Here, the top row shows projections of $s$ arranged by $\zeta$ and $\tau_{\rm a}$ values. In the purely solenoidal case, the overall density structure appears to be similar between the two runs, and this visual impression is consistent with the more detailed statistics that we present below. As we move toward the compressive runs, large low-density ``voids" are seen, which are most prominent in the $\zeta=0$ run. These changes are expected from previous studies, which show that the variance of $s$ increases with the compressive fraction \citep{Federrath08, Schmidt09, Konstandin12, Hopkins13a, Federrath13, Squire17}. 

What is unexpected from previous studies, however, is that in the simulations that include compressive modes, the fraction of the simulation volume occupied by the voids is larger in the $\tau_{\rm a} \approx 0.6 \tau_{\rm e}$ cases than in the $\tau_{\rm a} \approx 0.06 \tau_{\rm e}$ cases. This is most obvious in the $\zeta=0.0$ case and also in the slices of $s$ shown in the second row, which emphasize that the boundaries of these low-density regions are ringed by narrow high-density sheets and filaments.

The third row of this plot shows the divergence of the velocity field. Note that this Eulerian quantity is directly related to the Lagrangian change in $s$ measured by our tracer particles. This can be seen through a change of variables:
\be
\frac{d s}{d t} = \frac{1}{\rho} \left[ \frac{\partial \rho}{\partial t} + v_i \frac{\partial \rho}{\partial x_i} \right] = - \frac{\partial v_i}{\partial x_i},
\label{eq:dstoV}
\ee 
where the second equality makes use of the continuity equation (eq.\ \ref{eq:continuity}). Thus, this slice shows that the regions within the voids are filled with particles in which $\frac{ds}{dt} = -\nabla \cdot {\mathbf v}$ is moderately negative, while in the surrounding shocks, $\frac{ds}{dt}$ is strongly positive. Note that these shocks are not precisely aligned with the narrow high-density regions seen in the second row. Rather, they mark regions at the edges of the voids where the change in $\frac{ds}{dt}$ is high, and just behind these shocks, the density increases sharply.

The final row of this plot shows the divergence of the driven accelerations, which is zero in the solenoidal case. As this driving is purely on large scales, these features are much larger in scale than the shocks and filaments that dominate the denser structures in the simulations. On the other hand, they are similar to the voids seen in the large $\tau_{\rm a}$ simulations.

\subsection{Probability Distribution Functions}

\begin{figure}
\begin{center}
\includegraphics[width=0.47\textwidth]{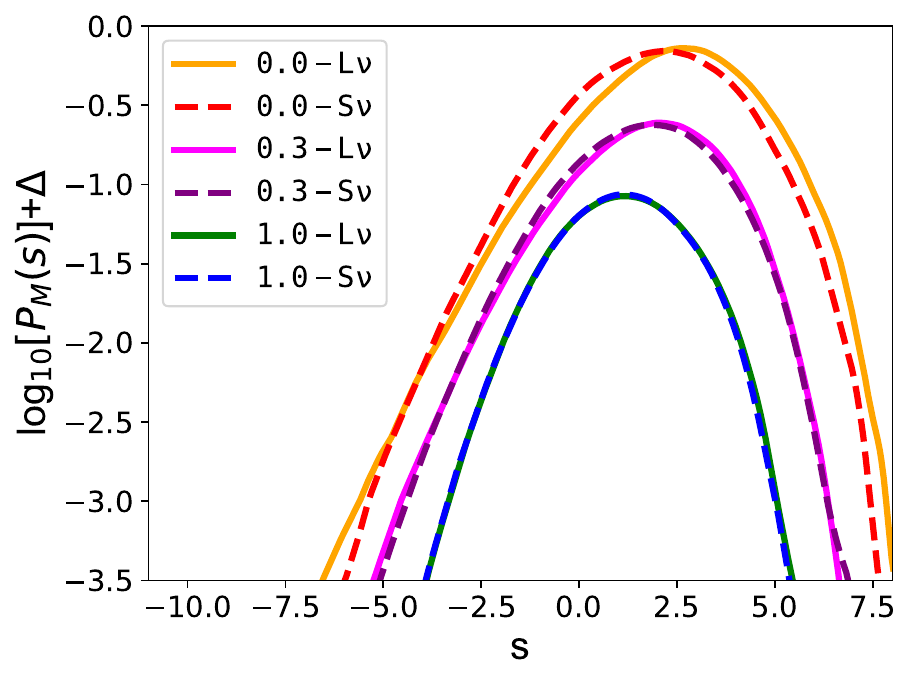}
\includegraphics[width=0.47\textwidth]{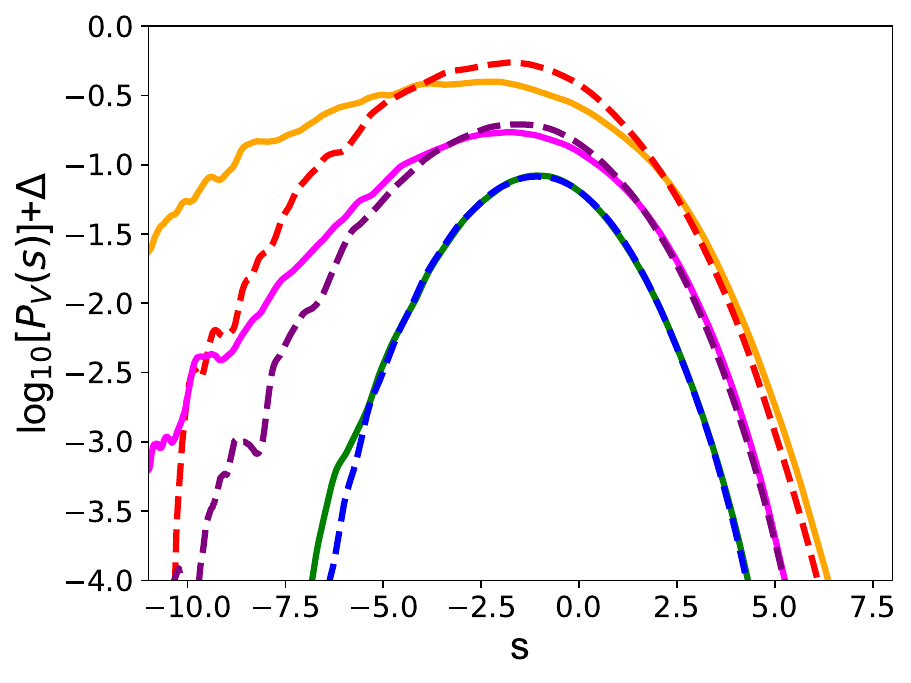}
\includegraphics[width=0.47\textwidth]{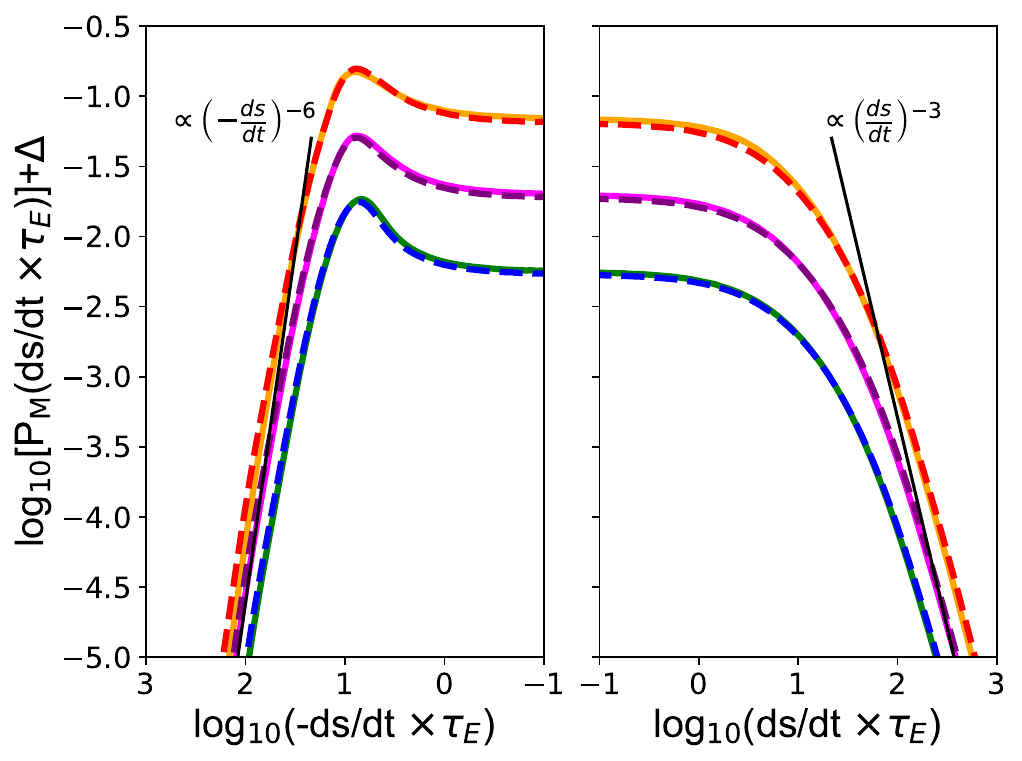}
\end{center}
\vspace{-0.2in}
\caption{{\em Top:} Mass-weighted PDF of $s$. As in Fig.\ \ref{fig:Ek}, the colored lines show $P_{\rm M}(s)$ for runs 0.0-L$\nu$ (orange solid), 0.0-s$\nu$ (red dashed), 0.3-L$\nu$ (magenta solid), 0.3-s$\nu$ (purple dashed), 1.0-L$\nu$ (green solid), and 1.0-s$\nu$ (blue dashed). For clarity, both fully-compressive, $\zeta=0$, runs are shifted upward by $\Delta = 0.5$, and both fully-compressive, $\zeta=1$, runs are shifted downward by $\Delta = 0.5$. {\em Center:} Volume-weighted PDF of $s$, with line styles and shifts as in the mass-weighted case. {\em Bottom:} The probability distribution of $\frac{ds}{dt}$. Lines are as in the top panel, and the black lines are $\propto (-\frac{ds}{dt})^{-6}$ on the left and $\propto (-\frac{ds}{dt})^{-3}$, illustrating the overall asymmetry in the tails of $P_{\rm M}(\frac{ds}{dt}\times \tau_{\rm e})$. The PDFs of the $\zeta=0$ and $\zeta=1$ runs are shifted for clarity, as in the upper panel.}
\label{fig:PS}
\end{figure}

In the top panel of Fig.\ \ref{fig:PS}, we show the mass-weighted probability distribution of $s$ for our viscous simulations as a function of forcing type and correlation time. We find that in the purely-solenoidal case ($\zeta=1$), the $P_{\rm M}(s)$ curves from the runs with different correlation times are extremely similar, even though in one case $\tau_{\rm a} \approx \tau_{\rm e}$ and in another $\tau_{\rm a}$ is less than 10\% of $\tau_{\rm e}$. This is consistent with the visual impression from Fig.\ \ref{fig:slice}.

Table \ref{tab:sigmas} quantifies the mean, variance, and skewness of these distributions, which are very similar between the two runs. In both cases, the distributions are nearly Gaussian, with a weak negative skewness ($\mu_{s,M} = \left< (s - \langle s \rangle)^3 \right>/\sigma_{s,M}^3$) due to shocks slowing down in the high-density regions, as described in detail in \cite{Paper1}. The similarity between the two runs is also evident in the volume-weighted PDF of $s$ shown in the central panel of Fig.\ \ref{fig:PS}, and the volume-weighted statistics in Table \ref{tab:sigmas}.

On the other hand, for the mixed driving case ($\zeta=0.3$), $P_{\rm M}(s)$ in the run 0.3-S$\nu$ with a correlation time of $\tau_{\rm a} =0.06 \tau_{\rm e}$ is notably different from the 0.3-L$\nu$ run with $\tau_{\rm a} =0.61 \tau_{\rm e}$. Although the variance is similar, Table \ref{tab:sigmas} shows that the skewness of $P_{\rm M}(s)$ in the 0.3-L$\nu$ run is significantly larger.

As the increase in negative skewness corresponds to an increase in the low-density tail, the differences between the two runs are much more apparent in the volume-weighted PDF shown in the central panels of this figure. In this case, $\sigma^2_{s,V}$ is $5.54$ in 0.3-L$\nu$, as compared to $4.10$ in the 0.3-S$\nu$ run, indicating that the volume-weighted distribution in turbulence that includes compressive driving is dependent on the driving correlation time.

Finally, for the purely-compressive case ($\zeta=0$), the variance of $P_{\rm M}(s)$ is similar between runs with different $\tau_{\rm a}$ values, but in this case, the skewness in the 0.0-L$\nu$ run is almost twice that in the 0.0-S$\nu$ run. For these runs, $P_{\rm V}(s)$ is completely different between the two runs, such that $\sigma^2_{s,V}$ is $9.07$ in 0.0-L$\nu$, as compared to $5.37$ in 0.0-S$\nu$. In other words, for compressively-driven turbulence, {\em the driving correlation time plays an essential role in setting properties of the density distribution.}

\begin{table*}[t]
 \centering
\hspace{-1.2in}
\begin{tabular}{lcccccccccccccc}
\toprule
 Name & $M_{\rm M}$ & $M_{\rm V}$ & 
 $\langle s \rangle$ & $\sigma^2_{s,M}$ & $\mu_{s,M}$ & $\langle s \rangle_{\rm V}$ & $\sigma^2_{s,V}$ & $\mu_{s,V}$ & $\langle \frac{ds}{dt} \rangle_M \tau_{\rm e}$ & $\sigma_{{ds}/{dt},M}^2 \tau_{\rm e}^2$ & $F_{{ds}/{dt}^+}$ & $\sigma_{{ds}/{dt}^+,M}^2/ \sigma_{{ds}/{dt},M}^{2}$\\
\midrule
1.0-L$\nu$ 	& 6.8 & 7.2 &1.10 & 2.10 & -0.12 & -1.13 & 2.33 & -0.08 & 0.034 & 1,900 & 0.25 & 0.91 \\
1.0-S$\nu$ 	 & 6.6 & 7.0 &1.09 & 2.07 & -0.14 & -1.12 & 2.29 & -0.01 & -0.025 & 2,040 & 0.25 & 0.91 \\
0.3-L$\nu$ 	& 6.2 & 7.3 &1.77 & 2.75 & -0.39 & -2.18 & 5.54 & -0.34 & -0.105 & 2,030 & 0.26 & 0.91\\
0.3-S$\nu$ 	 & 7.0 & 7.3 & 1.64 & 2.80 & -0.27 & -1.83 & 4.10 & -0.13 & 0.081 & 2,210 & 0.26 & 0.90 \\
0.0-L$\nu$ 	& 5.6 & 7.3 & 2.41 & 3.35 & -0.43 & -3.37 & 9.07 & -0.18 & 0.013 & 1,910 & 0.27 & 0.91 \\
0.0-S$\nu$ 	& 7.1 & 7.4 & 1.97 & 3.29 & -0.26 & -2.28 & 5.37 & -0.17 & 0.117 & 2,230 & 0.26 & 0.91\\
\bottomrule
\end{tabular}
\vspace{0.1in}
\caption{Properties of the simulated density distributions. Columns show the run name, the mass-weighted and volume-weighted Mach numbers ($M_{\rm M}$ and $M_{\rm V}$), the mean, variance, and skewness of $P_{\rm M}(s)$ as computed from the particles ($\langle s \rangle$, $\sigma^2_{s,M}$, and $\mu_{s,M}$), the mean, variance, and skewness of $P_{\rm V}(s)$ as computed from the grid ($\langle s \rangle_{\rm V}$, $\sigma^2_{s,V}$, and $\mu_{s,V}$), the mean and variance of $P_{\rm M}(\frac{ds}{dt})$ as computed from the particles ($\langle \frac{ds}{dt} \rangle_M \tau_{\rm e}$ and $\sigma_{{ds}/{dt},M}^2 \tau_{\rm e}^2$), the fraction of particles undergoing compressions ($F_{{ds}/{dt}^+}$), and the fraction of the variance in $\frac{ds}{dt}$ due to compressions ($\sigma_{{ds}/{dt}^+,M}^2/ \sigma_{{ds}/{dt},M}^{2}$).}
\label{tab:sigmas}
\end{table*}%

The bottom panel of Fig.\ \ref{fig:PS} shows the mass-weighted probability distribution of $\frac{ds}{dt}$ for all our simulations. Unlike the $s$ distribution, $P_{\rm M}\left(\frac{ds}{dt} \right)$ is similar across simulations, as quantified in Table \ref{tab:sigmas}. As in the solenoidal cases in \cite{Paper1}, we find that $P_{\rm M}(\frac{ds}{dt})$ is always asymmetric, with a peak to the left of $\frac{ds}{dt}=0$, indicating that most of the mass in the turbulent medium is slowly expanding. Also, as in \cite{Paper1}, we find that $P_{\rm M} \appropto(-\frac{ds}{dt})$ drops more sharply at negative values, with $P_{\rm M} \appropto(-\frac{ds}{dt})^{-6}$ for negative $\frac{ds}{dt}$ and $P_{\rm M} \appropto(\frac{ds}{dt})^{-3}$ for positive $\frac{ds}{dt}$. 

In \cite{Paper1}, we found that $\sigma_{{ds}/{dt},M}^2$ increased strongly with Mach number, indicating stronger shocks at higher $M_V$ values, and $\sigma_{{ds}/{dt},M}^2$ decreased strongly with viscosity, indicating broader shock fronts at higher $\nu_0$ values. Here we find that for fixed $M_{\rm V}$ and $\nu$ values, $\sigma_{{ds}/{dt},M}^2$ remains largely insensitive to the driving mechanism and correlation time. 

To maintain a steady state, $\frac{ds}{dt}$ must average to zero, and this holds true in all cases to within a few percent. If we integrate the right half of the histogram to compute the compressing mass fraction, $F_{{ds}/{dt}^+} \equiv \int_0^{\infty} d (\frac{ds}{dt}) P_{\rm M}(\frac{ds}{dt})$, we find that this is almost the same, $\approx 0.25$, in all cases. 

Likewise, the fraction of the variance that is due to compressions, which we label as $\sigma_{{ds}/{dt}^+,M}^2/ \sigma_{{ds}/{dt},M}^{2}$ in Table \ref{tab:sigmas}, is also the same, $\approx 0.9$, across the different runs. This indicates that strong compressions always dominate the variance of $\frac{ds}{dt}$ and that this compressive fraction is very similar in all cases. However, as we shall see below, this uniformity between runs breaks down if one considers the variance of $\frac{ds}{dt}$ in regions with different densities, and this provides insights into the origin of the strong $\tau_{\rm a}$ dependence of $P_{\rm V}(s)$ in the compressive cases.

\subsection{Conditional Averages}
\label{sec:conditional}

\begin{figure}[t]
\begin{center}
\includegraphics[width=0.47\textwidth]{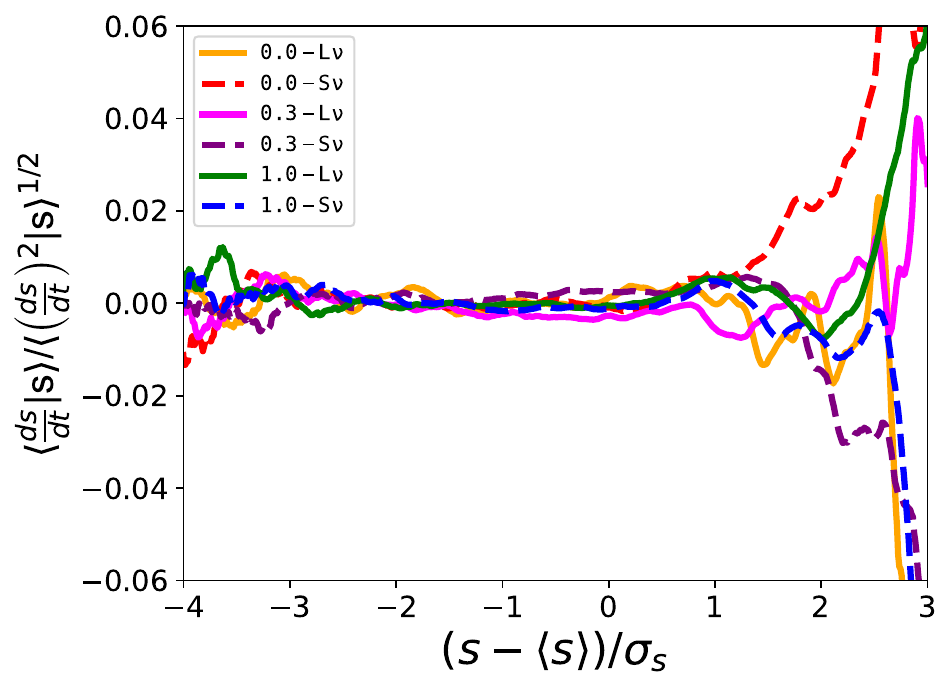}
\end{center}
\vspace{-0.2in}
\caption{Ratio of the average value of $\frac{ds}{dt}$ as a function of $s$ normalized by $\sigma_{{ds}/{dt},M}$. The lines are as above. This quantity should be zero in a steady state, and it thus serves as a test of the accuracy of our measurements \citep{Pan18}.}
\label{fig:Dsdt_norm}
\end{figure}

\begin{figure*}[t]
\begin{center}
\includegraphics[width=0.47\textwidth]{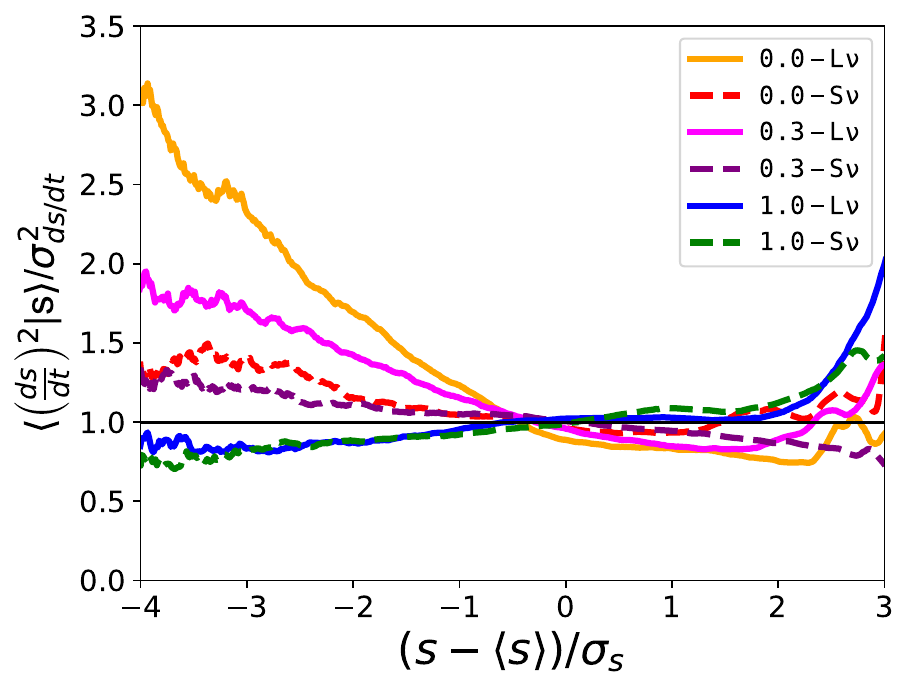} \qquad 
\includegraphics[width=0.47\textwidth]{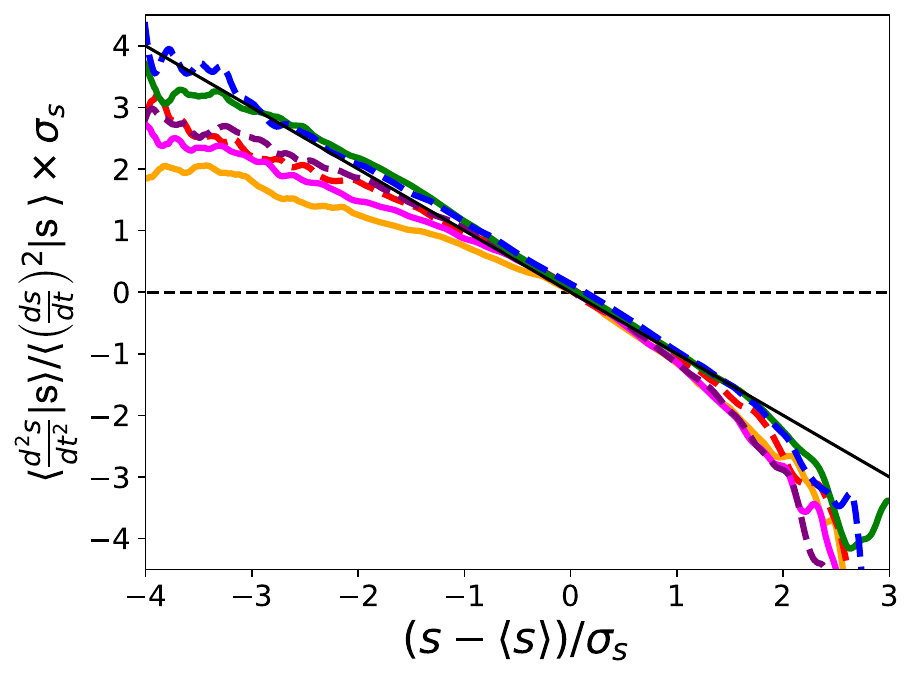}
\end{center}
\vspace{-0.2in}
\caption{{\em Left:} The average value of $\left< (\frac{ds}{dt})^2 | s \right>$ as a function of $s$ normalized by $\sigma_{{ds}/{dt},M}^2$. The lines are as above. The fully compressive $\zeta=0$ and partially compressive $\zeta=0.3$ runs driven with long correlation times (orange and magenta) show a much larger variance in $(\frac{ds}{dt})^2$ at low values of $s$ than similar runs with short correlation times (red and purple). These differences are not seen in the solenoidal runs. {\em Right:} The average value of the acceleration of the change in density $\left< \frac{d^2s}{dt^2} | s \right>$ as a function of $s$, normalized by $\left< (\frac{ds}{dt})^2 | s \right>/\sigma_{s,M}$. The line styles and colors are as in the previous figures.}
\label{fig:DsdtsqandR}
\end{figure*}

As discussed in \cite{Paper1}, in a steady state, $P_{\rm M}(s)$ can be directly related to the ensemble average of the time derivatives of $s$ conditioned on $s$ \citep{Pope93,Nordlund99,Pan19}.  Since our Lagrangian tracer particles track $s$  and its rate of change, we can readily compute these conditional derivatives.

The simplest such quantity is $\left< \frac{ds}{dt} | s \right>$, the average value of $\frac{ds}{dt}$ conditioned on $s$. In a steady state, $P_{\rm M}(s)$ is time-invariant, i.e., $\frac{d P_{\rm M}(s)}{dt} = 0$, which requires the net probability flux into and out of each $s$ bin to be zero \citep{Pan18}. Intuitively, the probability flux is given by $P_{\rm M}(s) \left< \frac{ds}{dt} | s \right> $, where $\left< \frac{ds}{dt} | s \right>$ is the average value of $\frac{ds}{dt}$ for particles within a given $s$ bin. Setting this flux to zero in a steady state leads to
\be
\left< \frac{ds}{dt} | s \right> = 0,
\label{eq:dsdtzero}
\ee
for all $s$. 

Fig.\ \ref{fig:Dsdt_norm} shows that in all simulations, the errors are below 1\% for $\left< s \right> -4 \sigma_{s,M} \leq s \leq \left< s \right> + 2 \sigma_{s,M}$. The errors are also less than 6\% in the $\left< s \right> + 2 \sigma_{s,M} \leq s \leq \left< s \right> + 3 \sigma_{s,M}$ a range in which much fewer particles are present due to the negative skewness of $P_{\rm M}(s)$. These results align with \cite{Paper1}, derived from \flash/ simulations \citep{Fryxell00}, confirming the accuracy of our approach using the \athenapk/ code.

To maintain a steady state, the time derivative of the probability flux must also be zero, $\frac{d^2 P_{\rm M}(s)}{dt^2} =0.$ This results in a second condition \citep{Pope93,Nordlund99,Pan19} that can be used to better understand the processes that lead to the underlying PDF:
\be
 P_{\rm M}(s) \left< \frac{d^2 s}{dt^2} | s \right> 
 -\frac{d}{ds} \left[ P_{\rm M}(s) \left< \left(\frac{ds}{dt} \right)^2 | s \right> \right] = 0. 
\label{eq:second}
\ee
Solving this equation for $P_{\rm M}(s)$ gives
\be
P_{\rm M}(s) \propto \frac{1}{\left< \left(\frac{ds}{dt} \right)^2 | s \right>} \exp \left[ \int_0^s ds' \frac{\left< \frac{d^2 s}{dt^2} | s' \right>}{\left< \left(\frac{ds}{dt} \right)^2 | s' \right>} \right],
\label{eq:psconditional}
\ee
which shows that the PDF of $s$ is fully determined by the conditional averages of $\left< \left(\frac{ds}{dt} \right)^2 | s \right>$ and $\left< \frac{ds^2}{d^2 t} | s \right>$.  This equation was confirmed in \cite{Paper1} and it is also confirmed in the right panel of Fig.\ \ref{fig:converge_ratiocheck} in the Appendix.

The left plot of Fig.\ \ref{fig:DsdtsqandR} shows $\left< \left(\frac{ds}{dt} \right)^2 | s \right>$ normalized by the variance of $\frac{ds}{dt}.$ At large $s$ values, a weak increase is observed in the solenoidal cases. As discussed in more detail in the Appendix, this is likely a numerical artifact, and it has no direct impact on the density distribution.

At $s$ values below $\left< s \right>,$ however, $\left< \left(\frac{ds}{dt} \right)^2 | s \right>$ shows a larger and gradual rise, which is strongly dependent on $\zeta$ and $\tau_{\rm a}.$ This quantifies the effect seen in Fig.\ \ref{fig:slice}, where large voids arise in the compressive-driving cases with small $\zeta$ and large $\tau_{\rm a}.$ In these regions, $\frac{ds}{dt}$ is moderately negative across most of the volume but highly positive at the edges, leading to high values of $\left(\frac{ds}{dt}\right)^2$.

The right panels of Fig.\ \ref{fig:DsdtsqandR} show the ratio of $\left< \frac{d^2s}{dt^2} | s \right>$ and $\left< \left(\frac{ds}{dt} \right)^2 | s \right>.$  This quantity, which appears in the exponential term in eq.\ (\ref{eq:second}) largely determines the shape of $P_{\rm M}(s)$. As discussed in \cite{Paper1}, this plot shows that shocks are systematically stronger in low-density regions and weaker in high-density regions, which leads to the approximately Gaussian form of $P_{\rm M}(s)$.  It also shows a downturn at small $s$ values, which is due to the increase of $\left< \left(\frac{ds}{dt} \right)^2 | s \right>$ shown in the left panels ($\left< \frac{d^2s}{dt^2} | s \right>$, which is not shown, is similar between the various runs).  Finally, all simulations show a noticeable downturn at large $s$ values. This corresponds to a strong drop in $\left< \left(\frac{ds}{dt} \right)^2 | s \right>$ as shocks are weakened as they move into the densest regions where the thermal pressure becomes comparable to the ram pressure. 

To understand the behavior of conditional averages and their dependence on $\zeta$ and $\tau_{\rm a}$, we make use of an equation derived in \cite{Pan19}:
\be
\frac{d^2s}{dt^2}=-\frac{d \left (\frac{\partial v_i}{\partial x_i}\right)}{dt} = \frac{\partial v_j}{\partial x_i} \frac {\partial v_i}{\partial x_j} + \frac {\partial \left(\rho^{-1} \frac{\partial p}{\partial x_i}\right)}{\partial x_i} - \frac {\partial\left(\rho^{-1} \frac{\partial \sigma_{ij}}{\partial x_j}\right)}{\partial x_i} - \frac{\partial a_i}{\partial x_i}.
\label{eqccel}
\ee
The effects of the nonlinear term, the pressure term, and the viscosity term on the right-hand side on the conditional average $\langle \frac{d^2s}{dt^2}|s\rangle$ have been extensively examined in Pan et al.\ (2019). In particular, the nonlinear term acts as a source driving the velocity divergence (or equivalently $\frac{ds}{dt}$), while the pressure term suppresses it, and the viscosity term exhibits more complex behavior.

For solenoidal driving, $\nabla \cdot {\mathbf a} =0$, meaning the force does not directly influence $\nabla \cdot {\mathbf v}$ or $\frac{ds}{dt}$. It is also likely that the driving acceleration does not have an indirect impact on the conditional averages, $\langle \left(\frac{ds}{dt}\right)^2|s\rangle$ and $\langle \frac{d^2s}{dt^2}|s\rangle,$ either. This follows because the first three terms on the right-hand side of eq.\ (\ref{eqccel}) all involve spatial derivatives and thus correspond to small scales that are insensitive to the pattern of large-scale driving. Indeed, the PDF of $s$ is essentially invariant with the correlation time of $a_i$ for solenoidal driving. 

\begin{figure}
\begin{center}
\includegraphics[width=0.47\textwidth]{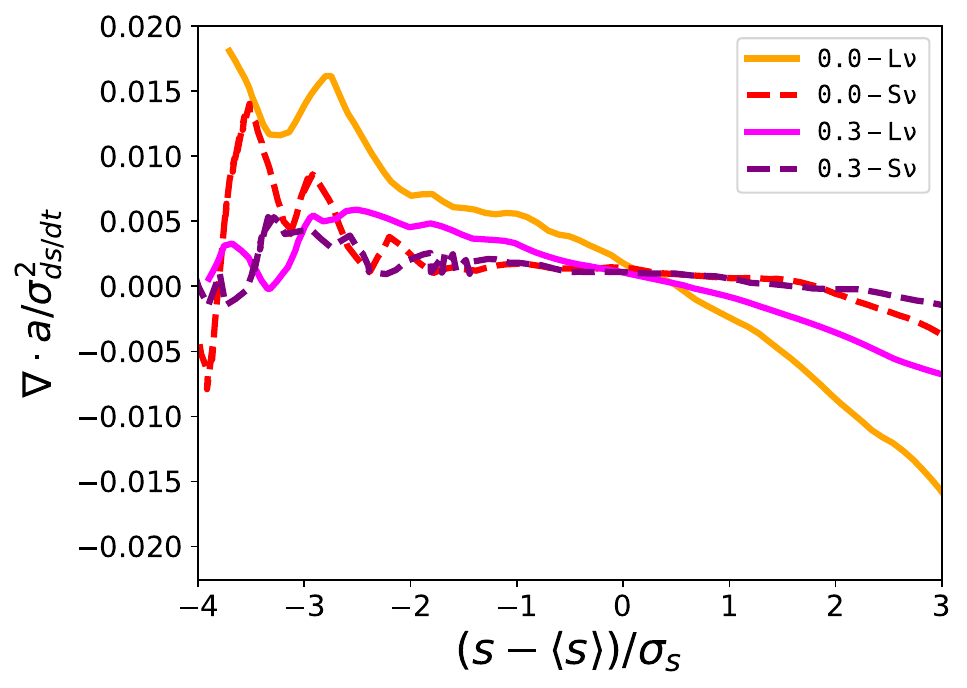} \qquad 
\end{center}
\vspace{-0.2in}
\caption{Divergence of the acceleration field, normalized to be directly comparable to the right panel of Fig.\ \ref{fig:DsdtsqandR}.}
\label{fig:dela}
\end{figure}

For compressive driving, the situation is different. In this case, $\nabla \cdot {\mathbf a}$ is nonzero and it acts as an external source for $\nabla \cdot {\mathbf u}$ or $\frac{ds}{dt}$. In Fig.\ \ref{fig:dela} we show $\langle \nabla \cdot {\mathbf a}|s\rangle$ normalized by $\left< \left(\frac{ds}{dt} \right)^2 | s \right>/\sigma_s.$ Comparing this with Fig.\ \ref{fig:DsdtsqandR} shows that $\langle \nabla \cdot {\mathbf a}|s\rangle$ is consistently much smaller than $\langle \frac{d^2s}{dt^2}|s\rangle$. 
The likely reason is that, as ${\bf a}$ is applied at large scales, eq.\ (\ref{eqccel}) suggests that its contribution to $\langle \frac{d^2s}{dt^2}|s\rangle$ is negligible in comparison to the first three terms on the right-hand side, which are all small-scale quantities. Also note that, in the special case where $a_i$ is white noise in time (i.e., a Wiener process), it is straightforward to see that $\langle \nabla \cdot {\mathbf a}|s\rangle =0$, meaning that the driving force does not directly contribute to $\langle \frac{d^2s}{dt^2}|s\rangle.$  

However, even though, the direct contribution of $\langle \nabla \cdot {\mathbf a}|s\rangle$ to $\langle \frac{d^2s}{dt^2}|s\rangle$ is negligible, $\nabla \cdot {\mathbf a}$ may impact the conditional variance, $\langle \left(\frac{ds}{dt}\right)^2|s\rangle.$ eq.\ (\ref{eqccel}) shows that, despite the small magnitude of $\nabla \cdot {\mathbf a}$, it provides a seed for the evolution of $\frac{ds}{dt}$. 
This seed can be amplified by the nonlinear term in eq. (\ref{eqccel}), and may considerably affect $\langle \left(\frac{ds}{dt}\right)^2|s\rangle$ if the amplification time is sufficiently long.

The impact of $\nabla \cdot {\mathbf a}$ on $\langle \left(\frac{ds}{dt}\right)^2|s\rangle$ through nonlinear amplification depends on both the driving correlation time and the evolution timescale of $\frac{ds}{dt}$ along a Lagrangian trajectory. Intuitively, the slower $\frac{ds}{dt}$ evolves, the easier it is for the contribution of $\nabla \cdot {\mathbf a}$ to grow. Also, the larger $\tau_{\rm a}$, the more coherent $\nabla \cdot {\mathbf a}$ is in time and the easier it may accumulate and be amplified by the nonlinear term. This means that its contribution to $\langle (\frac{ds}{dt})^2|s\rangle$ would increase with increasing $\tau_{\rm a}$.

The evolution timescale of $\frac{ds}{dt}$ in low-density regions with smaller $s$ is expected to be larger due to their larger width and smoother configuration, while in the high-density regions, the evolution of $\frac{ds}{dt}$ is expected to be faster. Thus, the impact of $\nabla \cdot {\mathbf a}$ on $\langle (\frac{ds}{dt})^2|s\rangle$ grows faster and persists longer at smaller $s$. This explains the result shown in the left panel of Fig.\ \ref{fig:DsdtsqandR}, for the strongly compressive runs with $\zeta=0$ and $\zeta=0.3.$ Here we see that, as the correlation time of $a_i$ increases, $\langle (\frac{ds}{dt})^2|s\rangle$ becomes significantly larger at small $s$, where $\frac{ds}{dt}$ evolves slowly, but remains almost invariant at large $s$, where the evolution timescale of $\frac{ds}{dt}$ is short. 

In the right panels of Fig.\ \ref{fig:DsdtsqandR}, we see that the increases in $\langle \left(\frac{ds}{dt}\right)^2|s\rangle$ at low $s$ values in the compressive cases with long driving correlation times causes a decrease in the ratio of
$\langle \frac{d^2s}{dt^2} |s\rangle$ and $\langle \left(\frac{ds}{dt}\right)^2|s\rangle.$ This in turn leads to a more negatively-skewed $P_{\rm M}(s).$ Or in other words, for compressively-driven turbulence, increasing the correlation time increases the variance of the velocity divergence in low-density regions, and this leads to a more negatively skewed in $P_{\rm M}(s)$ and broader $P_{\rm V}(s).$

\subsection{Correlation Between Density and Compressive Driving}

\begin{figure*}[t]
\begin{center}
\includegraphics[width=0.98\textwidth]{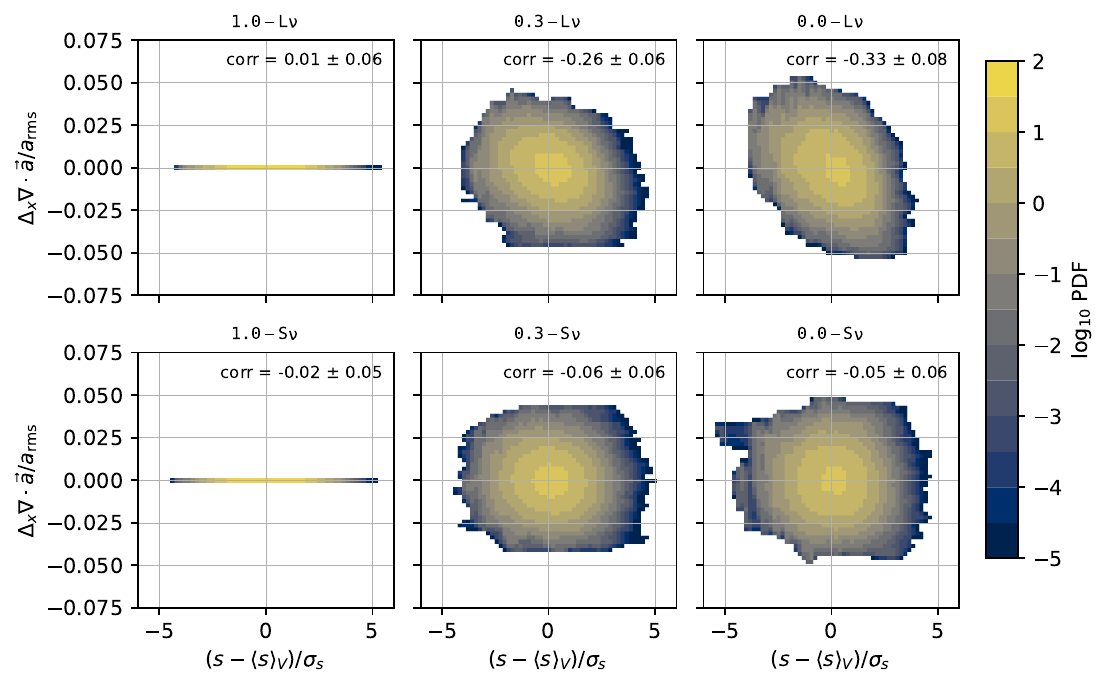}
\end{center}
\vspace{-0.2in}
\caption{Two-dimensional volume-weighted PDF of normalized log density $s-\left < s \right >_v$ and normalized divergence of the acceleration field $\Delta_x \nabla \cdot \mathbf a / a_{\rm rms}.$ Each panel provides the correlation coefficient between the two quantities. Only for large $\tau_{\rm a}$ values in combination with compressive driving, $\Delta_x \nabla \cdot \mathbf a / a_{\rm rms}$ are negatively correlated with $s,$ indicating that the driven accelerations lead directly to regions of low density. In all other cases, there is no significant correlation.}
 \label{fig:svsdiva}
\end{figure*}

To better understand the increase in $\left< \left(\frac{ds}{dt} \right)^2 | s \right>$ at low $s$ in compressive, large $\tau_{\rm a}$ runs, we computed the 2D volume weighted PDF of log density $s-\left < s \right >_v$ and the divergence of the driven acceleration field $ \nabla \cdot \mathbf a$ as shown in Fig.~\ref{fig:svsdiva}. Note that, 
{\bf a} is added on large scales to sustain the overall level of turbulence. Thus in the $\zeta = 1$ runs, this divergence is zero.\footnote{Given that the driving is defined in spectral space, numerical noise exists for $\nabla \cdot \mathbf{a}$ calculated using cell-centered finite differences in real space.} This means that these distributions are equivalent to the volume-weighted $\zeta=1$ histograms of $s$ shown in the central panel of Fig.\ \ref{fig:PS}, which are almost identical between $\tau_{\rm a}= 0.63 \tau_{\rm e}$ and $\tau_{\rm a}= 0.07 \tau_{\rm e}$ runs.

For $\zeta < 1$ runs, on the other hand, changing $\tau_{\rm a}$ both increases the overall negative skewness of distribution and introduces a significant correlation between $s$ and $\nabla \cdot a.$ This is quantified by the Pearson product-moment correlation coefficient shown in each panel.
 
Note that ${\mathbf a}$ is added to the simulation according to eqs.\ (\ref{eq:driving}) and (\ref{eq:driving2}), and it is not affected by any properties of the medium. Thus the strong correlation between $s$ and $ \nabla \cdot \mathbf a$ implies that the large void regions seen in the large $\tau_{\rm a}$ runs {\rm are directly caused by the driving accelerations.} 

These are regions in which $\nabla \cdot \mathbf a$ is positive and, because $\tau_{\rm a}$ is large, accelerated expansions are sustained over a significant period of time. The result is large, underdense expanding regions, corresponding to a moderately negative $\frac{ds}{dt}$ over the majority of the volume, which are ringed by a shock of swept-up material, corresponding to a large $\frac{ds}{dt}$ near the boundaries. Conversely, if $\tau_{\rm a},$ is small, the accelerations are not sustained long enough to build up significant expansions, meaning that $ \nabla \cdot \mathbf a$ and $s$ remain largely uncorrelated, and the medium is unable to build up the voids.

A potential relationship between the correlation time in the driving and the low density tail of the density PDF has already been commented on by \citet{Konstandin12} but not explored in more detail. Moreover, \citet{Alvelius1999} showed analytically that there exists a velocity-force correlation in cases in which the correlation time of the forcing is not small enough (compared to the turbulence timescales).

\section{Discussion: Driving Correlation Time in a Supernova Driven Interstellar Medium}
\label{sec:discussion}
The strong influence of the driving correlation time on the density structure of compressively-forced, supersonic turbulence raises the question of which values of $\tau_{\rm a}/\tau_{\rm e}$ are most likely in astrophysical systems. One key example of such a system is supernova-driven turbulence in the interstellar medium \citep[e.g.,][]{Beck96,Elmegreen04,Hennebelle14,Walch15,Martizzi15,Padoan16,Girichidis16,Chamandy20,Bacchini20}. Capturing the full structure of the ISM requires detailed models that include chemical reactions \cite[e.g.,][]{vanDishoeck98, Glover12, Ramirez13}, magnetic fields \citep[e.g.,][]{Zweibel95, Li15, Xu19, Seifried20,Kim21}, and radiative transfer \citep[e.g.,][]{Kim18,Schneider20,Grudic21,Ostriker22}.

However, we can use a simple analytic model to provide us with a rough estimate of the approximate range of $\tau_{\rm a}/\tau_{\rm e}$ likely to be encountered. \cite{Blondin98} calculated the radius at which supernovae become momentum conserving (i.e., enter the snowplow phase) in a uniform medium containing solar metallicity gas \citep{Sutherland93} as $R_{\rm mom} \approx 20 \, {\rm pc} \, E_{51}^{5/17} n_0^{-7/17},$ where $E_{51}$ is the energy per supernova in units of 10$^{51}$ ergs and $n_0$ is the number density of 
the medium in units of cm$^{-3},$ which is related to the mass density as $\rho = 0.0145 M_\odot {\rm pc}^{-3} n_0.$ At this radius the total net momentum of the supernova remnant is $P_{\rm mom} \approx 3 \times 10^5 \, M_\odot \, {\rm km} \, {\rm s^{-1}} \, E_{51}^{16/17} n_0^{-2/17}$ estimated from the total swept up mass and blast velocity~\citep{Blondin98}.

We can use this expression to estimate the turbulent properties of the medium by balancing the energy input from supernovae with energy losses through the turbulent cascade \citep[see also][]{Martizzi15,MLi15}. This gives
\be
f_{\rm SN} \frac{1}{2} P_{\rm mom} v \dot n_{\rm SN} = \frac{1}{2} \rho \frac{\sigma^3}{L} ,
\ee
where $f_{\rm SN}$ is a constant of order unity, $v$ is the expansion velocity of the remnant, $\sigma$ is the velocity dispersion of the turbulence and $L$ is the driving scale of the turbulence. This means that the driving scale goes as
\be
L = \frac{\rho \sigma^3}{ P_{\rm mom} v f_{\rm SN} \dot n_{\rm SN}}.
\ee
In an isotropic medium, $L$ is likely be approximately the scale at which the velocity of the supernova is equal to the turbulent velocity of the medium, which is given by
\be
V(L) = \frac{ 3 P_{\rm mom}}{ 4 \pi \rho L^3}.
\label{eq:VofL}
\ee
Setting $V=\sigma$ and defining $\dot n_{\rm SN,12}$ as the number of supernovae per $10^{12}$ years per pc$^3$
gives
\be
\sigma = \left[\frac{3 P_{\rm mom}^4 f_{\rm SN}^3 \dot n_{\rm SN}^3} {4 \pi \rho^4} \right]^{1/7} 
= 33 \, {\rm km \, s}^{-1} \frac{ (f_{\rm SN} \, \dot n_{\rm SN,12})^{3/7} \, E_{\rm 51}^{64/119}}{n_0^{76/119}},
\label{eq:sigma}
\ee
and
\be
L = \left[ \frac{9 P_{\rm mom}}{f_{\rm SN} \, (4 \pi)^2 \, \rho \, \dot n_{\rm SN}} \right]^{1/7} 
= 45 \, {\rm pc}\, \frac{ ( f_{\rm SN} \ \dot n_{\rm SN, 12})^{-1/7} \, E_{\rm 51}^{16/119}}{n_0^{19/119}},
\label{eq:scalelength}
\ee
such that $L^3 \sigma = 3 P_{\rm mom}/(4 \pi \rho).$ These equations can also be adapted for clustered supernovae by interpreting $E_{\rm 51}$ as the energy per grouping of supernovae and $n_{\rm SN}$ as the number density of such groupings.

To compute the correlation time we can ask how long it takes for a supernova to expand to the driving scale. If this occurs entirely in the momentum-conserving regime, then
\be
\tau_{\rm a} = \frac{1}{4} L^4 \left[\frac{4 \pi \rho}{3P_{\rm mom}} \right] = \frac{1}{4} \tau_{\rm e},
\ee 
where $\tau_{\rm e} = L./\sigma.$ 
So in cases in which the radius at which the typical supernova becomes momentum conserving, $R_{\rm mom}$ is much less than the driving scale $L,$ the natural value of $\tau_{\rm a}/\tau_{\rm e} \approx 0.25,$ between the two cases considered in our simulations. 
 
On the other hand, if the radius at which the typical supernova becomes momentum conserving is significant, $R_{\rm mom} \approx L,$ then $\tau_{\rm corr}$ is shorter, meaning that $\tau_{\rm a}/\tau_{\rm e} < 0.25.$ This would occur at lower densities, higher supernova rates, or in cases with significant clustering of supernovae. Finally, if $R_{\rm mom} > L$, the SN will overlap in the energy-conserving phase, leading to blow-out. This would result in an outflowing galaxy, instead of a turbulent ISM.
 
Together these estimates suggest that conditions in the ISM are often likely to lead to driving correlation times less than $\tau_{\rm e},$ placing them between the two limits studied here. This underscores the need to explore the relationship between driving correlation timescales in more realistic models, incorporating a detailed description of the diverse physical processes at work in the ISM.

\section{Conclusions}
\label{sec:conclusion}

Supersonic turbulence plays a key role in astrophysical systems, from planetary to circumgalactic scales. While many studies have examined how density distributions depend on magnetic field strength, equation of state, Mach number, and driving mechanism, the role of the driving correlation time, $\tau_{\rm a},$ has received less attention.

In this work, we explored the impact of this timescale on a set of hydrodynamical simulations that spanned both compressive and solenoidal driving. For solenoidal driving, we found that varying $\tau_{\rm a}$ from values approximately equal to the eddy turnover time, $\tau_{\rm e},$ to values $0.1 \tau_{\rm e}$ had no significant effect on the density distribution, as quantified by the mass-weighted PDF of $s \equiv \ln (\rho/\rho_0)$. In contrast, for compressive driving, we found that the density distribution strongly depends on $\tau_{\rm a},$ with a much narrower distribution seen when $\tau_{\rm a} \ll \tau_{\rm e}.$ Furthermore, by including a set of tracer particles in our simulations, we tracked the Lagrangian evolution of density fluctuations and identified the mechanism behind this dependence.

Our main results can be summarized as follows:

\begin{itemize}

\setlength\itemsep{0em}

 \item In compressively-driven turbulence, the driving correlation time significantly influences the formation of large, low-density voids. When the driving correlation time is long ($\tau_{\rm a} \approx \tau_{\rm e}$), the medium is observed to host extensive underdense regions, which are surrounded by shocks with large density contrasts. These voids are much less prominent when the driving correlation time is short ($\tau_{\rm a} \approx 0.1 \tau_{\rm e}$), and completely absent in cases with solenoidal driving.
 
 \item The voids observed in compressively-driven turbulence are reflected in the mass-weighted probability distribution function of $s$, which depends strongly on $\tau_{\rm a}$. For long driving correlation times, $P_{\rm M}(s)$ becomes broader and more negatively skewed, while for short driving correlation times, it is narrower and more symmetric. These trends are weaker in cases in which turbulence is driven by a mixture of compressive and solenoidal modes, and in the purely solenoidal case, $P_{\rm M}(s)$ is nearly Gaussian and largely insensitive to $\tau_{\rm a}.$

 \item The variance of $\frac{ds}{dt}$ is always dominated by shocks. However, the conditional average $\left< \left(\frac{ds}{dt}\right)^2|s \right>$ reveals that the driving correlation time affects the evolution of density most strongly in low-density regions. This likely arises from the impact of $\nabla \cdot {\mathbf a}$ on $\langle (\frac{ds}{dt})^2|s\rangle$ through nonlinear amplification. Intuitively, the larger $\tau_{\rm a}$, the more coherent $\nabla \cdot {\mathbf a}$ is in time and the easier it may accumulate and be amplified. This leads to a larger variance in $\left< \left(\frac{ds}{dt}\right)^2|s \right>$ in low $s$ regions, which are expected to evolve more slowly and hence be more easily affected, and this larger variance leads to a more skewed and broader $P_{\rm M}(s)$. 

 \item The large variance of $\frac{ds}{dt}$ is consistent with the large underdense regions in the large $\tau_{\rm e}$ compressively-driven simulations. In these simulations, sustained accelerated expansions lead to the creation of slowly expanding voids that are bounded by shocks with large density contrasts, and this results in an overall high value of $\left(\frac{ds}{dt} \right)^2$ at low $s.$ These voids are directly linked to the divergence of the driving acceleration field, $\nabla \cdot \mathbf{a}$, which remains positively correlated with low-density regions over extended periods.

\end{itemize}

These results have important implications for astrophysical systems in which compressively-driven turbulence plays a key role, such as the interstellar medium and molecular clouds. While the energy injection processes are naturally more complex, the driving correlation time in these environments is often determined by supernovae, which can have correlation times significantly shorter than the eddy turnover time. This indicates that the density structure  will depend not only on the Mach number and driving mechanism but also on the overall driving correlation time, with strong implications for numerous astrophysical systems.

\section*{Acknowledgements}

PG acknowledges the support by the DFG grant SCHW 1358/5-1 within the priority program SPP 1992.
ES was supported by NASA grant 80NSSC22K1265 and the work includes simulations run on NASA High-End Computing resources. L.P. acknowledges financial support from NSFC under grant No.\ 11973098 and No.\ 12373072. M.B. acknowledges support from the Deutsche Forschungsgemeinschaft under Germany's Excellence Strategy - EXC 2121 "Quantum Universe" - 390833306 and from the BMBF ErUM-Pro grant 05A2023. The authors gratefully acknowledge the Gauss Centre for Supercomputing e.V. for funding this project by providing computing time through the John von Neumann Institute for Computing (NIC) on the GCS Supercomputer JUWELS at J\"ulich Supercomputing Centre (JSC).

\bibliographystyle{aasjournal}
\bibliography{Turb.bib}
\vspace{0.2in}

\section{Appendix}

\begin{figure}[t]
\begin{center}
 \includegraphics[width=0.45\textwidth]{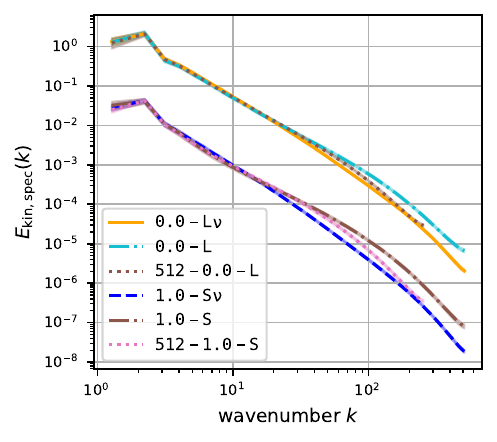}
 \includegraphics[width=0.47\textwidth]{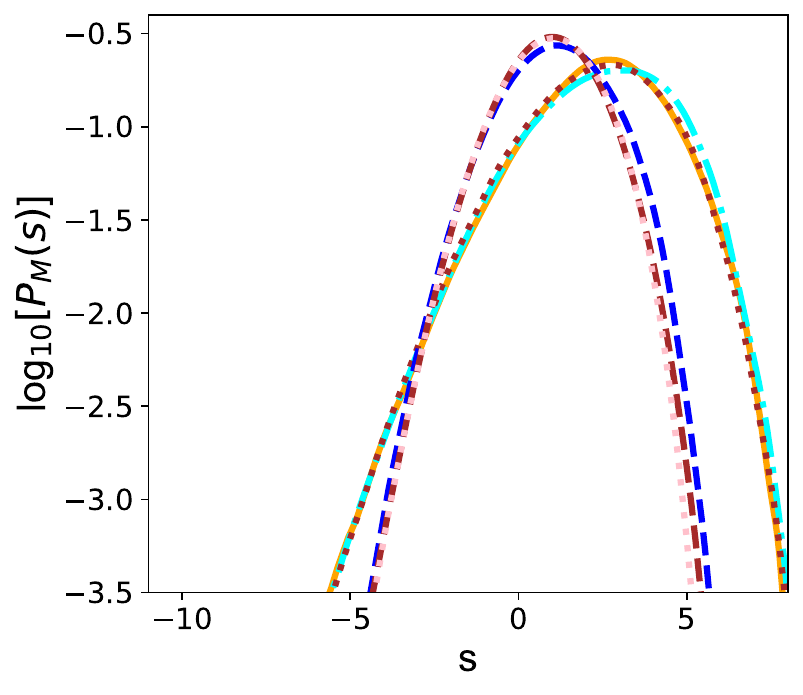}
\end{center}
\vspace{-0.2in}
\caption{Left: Specific kinetic energy spectra comparing different numerics ($1024^3$ DNS, $1024^3$ ILES, and $512^3$ ILES)
 at the extreme points in our parameter space, i.e., for $\zeta=0.0$ and long $\tau\mathrm{a}$, and $\zeta = 1.0$ and
 short $\tau\mathrm{a}$. 
The $\zeta=0.0$ spectra are vertically offset for better visibility.
Right: Mass-weighted PDFs of $s$ for the same simulations, with linestyles and colors as in the left panels.}
\label{fig:converge}
\end{figure}

To study the impact of numerical effects, in Fig.\ \ref{fig:converge} we compare two key quantities from our fiducial $1024^3$ simulations with corresponding ILES simulations with $1024^3$ and $512^3$ cells. As described in \S2.2, the explicit viscosity in the fiducial simulation matches the effective viscosity of the $512^3$ ILES runs. Here we focus on the most extreme cases: purely compressive runs with long driving correlation times and purely solenoidal runs with short correlation times.

The left panel compares kinetic energy spectra across these six simulations. Like in Fig.\ \ref{fig:Ek}, the spectra all exhibit similar slopes of $\approx 2$ in the inertial range. For $\zeta=0$ and $\tau_{\rm a} \approx \tau_{\rm e}$, the $1024^3$ run with explicit viscosity and the $512^3$ ILES run closely match, with a sharp turndown around $k \approx 128.$ This is consistent with their similar $\nu_{\rm eff}$ values of $4.9 \times 10^{-4}$ and $4.2 \times 10^{-4}$ (see Table \ref{tab:runs}). In the solenoidal ($\zeta=1$) case, on the other hand, the turndown in the $512^3$ ILES run occurs at $k \approx 30,$ while the turndown in the fiducial case occurs at $k \approx 20$ due to the stronger effects of shear viscosity. Again, this is reflected in their relative effective viscosities of $6.0 \times 10^{-4}$ and $10.2 \times 10^{-4}$, respectively.

\begin{table*}[t]
 \centering
\hspace{-1.2in}
\begin{tabular}{lcccccccccccccc}
\toprule
 Name & $M_{\rm M}$ & $M_{\rm V}$ & 
 $\langle s \rangle_M$ & $\sigma^2_{s,M}$ & $\mu_{s,M}$ & $\langle s \rangle_{\rm V}$ & $\sigma^2_{s,V}$ & $\mu_{s,V}$ & $\langle \frac{ds}{dt} \rangle \tau_{\rm e}$ & $\sigma_{{ds}/{dt},M}^2 \tau_{\rm e}^2$ & $F_{{ds}/{dt}^+}$ & $\sigma_{{ds}/{dt}^+,M}^2/ \sigma_{{ds}/{dt},M}^{2}$\\
\midrule
1.0-L 	& 6.8 & 7.1 &0.92 & 1.72 & -0.12 & -1.01 & 2.24 & -0.29 & -0.004 & 4,330 & 0.35 & 0.82 \\
1.0-S	 & 6.8 & 7.0 &0.92 & 1.74 & -0.13 & -1.01 & 2.19 & -0.24 & 0.054 & 4,300 & 0.34 & 0.82 \\
0.3-L		& 6.2 & 7.4 &1.81 & 2.86 & -0.34 & -2.25 & 5.69 & -0.26 & -0.001 & 4,210 & 0.34 & 0.83 \\
0.3-S	 & 7.1 & 7.4 & 1.42 & 2.46 & -0.26 & -1.62 & 3.67 & -0.17 & -0.033 & 4,610 & 0.34 & 0.83 \\
0.0-L		& 5.8 & 7.2 & 2.53 & 3.74 & -0.42 & -3.35 & 8.95 & -0.22 & 0.065 & 4,110 & 0.34 & 0.84 \\
0.0-S	& 7.1 & 7.3 & 1.34 & 2.80 & -0.26 & -1.89 & 4.40 & -0.23 & 0.049 & 4,790 & 0.34 & 0.84 \\[0.5em]
512-1.0-L & 7.1 & 6.9 & 0.94 & 1.75 & -0.16 & -1.01 & 2.18 & -0.20 & -0.002 & 2,660 & 0.34 & 0.83 \\
512-1.0-S & 6.9 & 6.6 & 0.93 & 1.73 & -0.18 & -0.99 & 2.10 & -0.13 & 0.016 & 1,570 & 0.34 & 0.83 \\
512-0.0-L & 7.1 & 5.9 & 2.34 & 3.53 & -0.37 & -3.00 & 8.05 & -0.30 & 0.055 & 1,650 & 0.34 & 0.85 \\
512-0.0-S & 7.5 & 7.4 &1.71 & 2.94 & -0.23 &-1.92 & 4.59 & -0.28 & 0.006 & 2,690 & 0.33 & 0.82 \\
\bottomrule
\end{tabular}
\vspace{0.2in}
\caption{Properties of the simulated density distributions. Columns show the run name, the mass-weighted and volume-weighted Mach numbers ($M_{\rm M}$ and $M_{\rm V}$), the mean, variance, and skewness of $P_{\rm M}(s)$ as computed from the particles ($\langle s \rangle,$ $\sigma^2_{s,M},$ and $\mu_{s,M}$), the mean, variance, and skewness of $P_{\rm V}(s)$ as computed from the grid ($\langle s \rangle_{\rm V},$ $\sigma^2_{s,V},$ and $\mu_{s,V}$ ) the mean and variance of $P_{\rm V}(\frac{ds}{dt})$ as computed from the particles ($\langle \frac{ds}{dt} \rangle \tau_{\rm e}$ and $\sigma_{{ds}/{dt},M}^2 \tau_{\rm e}^2$), the fraction of particles undergoing compressions ($F_{{ds}/{dt}^+}$), and the fraction of the variance in $\frac{ds}{dt}$ due to compressions ($\sigma_{{ds}/{dt}^+,M}^2/ \sigma_{{ds}/{dt},M}^{2}$).}
\label{tab:converge}
\end{table*}

\begin{figure}[t]
\begin{center}
 \includegraphics[width=0.47\textwidth]{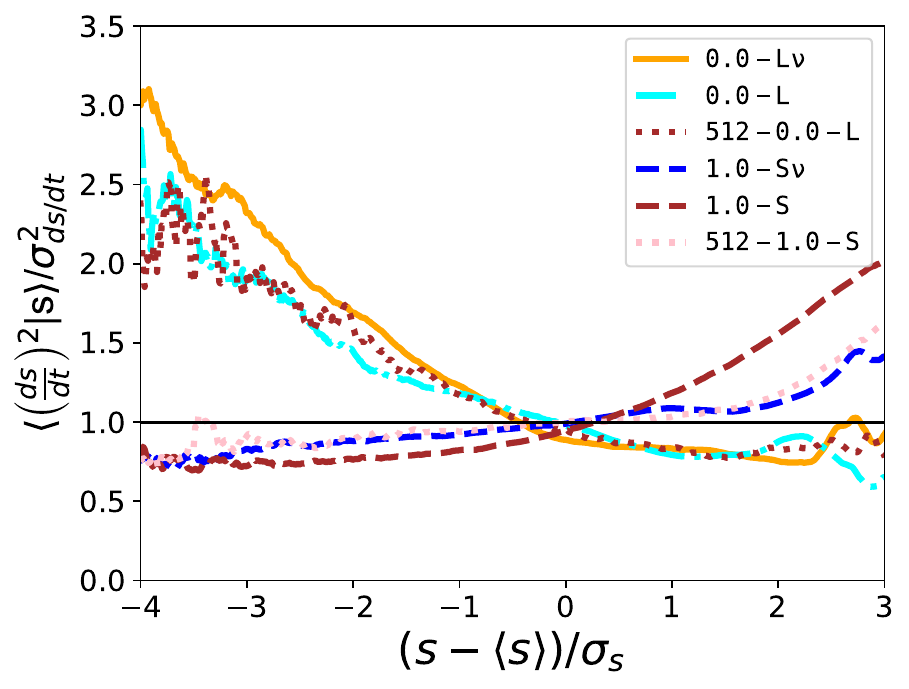}
 \includegraphics[width=0.47\textwidth]{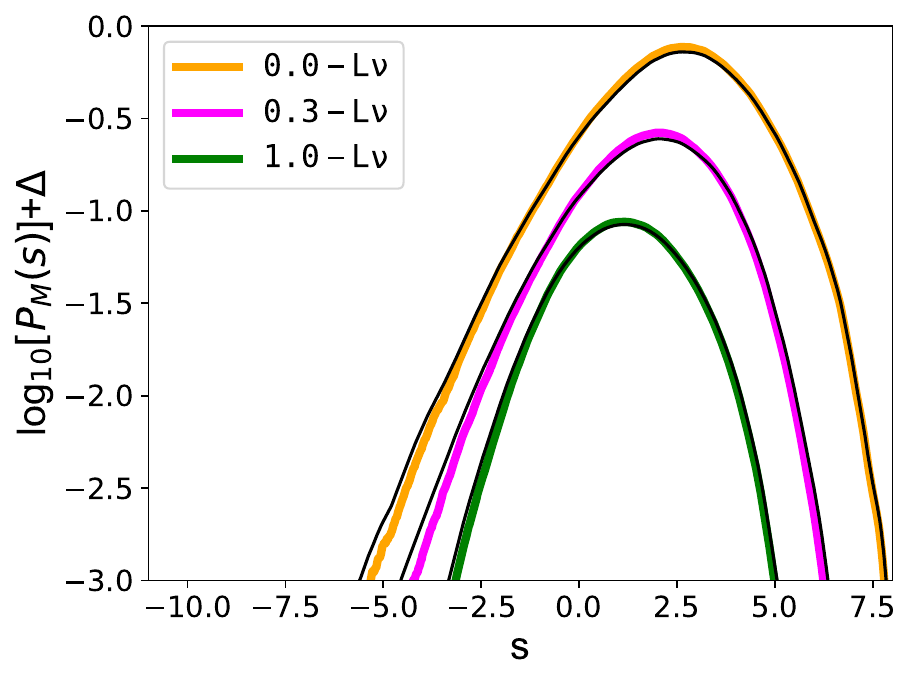}
\end{center}
\vspace{-0.2in}
\caption{Left: The average value of $\left< (\frac{ds}{dt})^2 | s \right>$ conditional on $s$ normalized by $\sigma_{{ds}/{dt},M}^2$ comparing different numerics ($1024^3$ DNS, $1024^3$ ILES, and $512^3$ ILES) at the extreme points in our parameter space. Right: Predictions for the mass-weighted $P_M(s)$ from our conditional averages for the fiducial simulations in our study with long $\tau_a$ values. The colored lines show $P_M(s)$ as computed from eq.\ (\ref{eq:psconditional}), while the black lines show $P_M(s)$ directly measured from the simulations with long driving times.}
\label{fig:converge_ratiocheck}.
\end{figure}

The right panel of Fig.\ \ref{fig:converge} shows the mass-weighted PDF of each of these simulations, as measured from the Lagrangian tracer particles, demonstrating that these distributions are independent of explicit viscosity or resolution. This is consistent with the mass-weighted and volume-weighted values of $\left<s \right>$, $\sigma^2_{\rm s}$, and $\mu_{\rm s}$ in Tables \ref{tab:sigmas} and \ref{tab:converge}, which show close agreement for cases with the same $\zeta$ and $\tau_{\rm a}$ values. These tables also include statistics from additional runs, enabling a broader comparison across the parameter space.

Table \ref{tab:converge} also includes statistics of $P_M\left(\frac{ds}{dt}\right)$. Here we see that in all cases $\langle \frac{ds}{dt} \rangle_M \tau_{\rm e}$ is small, consistent with a steady-state configuration. As discussed above, the variance $\sigma_{{ds}/{dt},M}^2 \tau_{\rm e}^2$ is set by the Mach number and the width of the shocks. Consequently, it is similar between the fiducial $1024^3$ simulations with explicit viscosity and the $512^3$ ILES simulations, but it is approximately twice as large in the $1024^3$ ILES case, indicating narrower shocks.

In the left panel of Fig.\ \ref{fig:converge_ratiocheck}, we show the normalized conditional average of $\left< (\frac{ds}{dt})^2 | s \right>$ for purely compressive runs with long driving correlation times and purely solenoidal runs with short driving correlation times. At low densities, these runs consistently exhibit features seen in the fiducial runs. In the solenoidal case $\left< (\frac{ds}{dt})^2 | s \right>$ is roughly constant below $\left<s \right> $ and for the compressive case there is a strong increase in $\left< (\frac{ds}{dt})^2 | s \right>$ in the lowest-density regions. This corresponds to expanding voids surrounded by shocks of swept-up material, and it results in a broad $P_V(s)$.

At high densities, the compressive runs yield similar results, while the solenoidal runs exhibit a weak increase with increasing $s.$ In this case, the fiducial $1024^3$ and $512^3$ ILES show a similar mild rise with increasing $s,$ while the rise is somewhat larger in the $1024^3$ run. Note however that these features have only a minor effect on $\left< \frac{d^2s}{dt^2} | s \right>/\left< (\frac{ds}{dt})^2 | s \right>$ (not shown here) which is dominated by a significant downturn at large $s$ values. This ratio largely sets the shape of $P_M(s)$ (see eq. \ref{eq:psconditional}) and it is very similar across runs.

The mass fraction of the gas undergoing compression at any given time is somewhat higher in the ILES runs ($F_{{ds}/{dt}^+} \approx 0.35$) compared to the fiducial runs ($F_{{ds}/{dt}^+} \approx 0.25$). Conversely, the fraction of variance due to compressions is lower in the ILES runs ($\sigma_{{ds}/{dt}^+,M}^2/ \sigma_{{ds}/{dt},M}^{2} \approx 0.8$) than in the viscous runs ($\sigma_{{ds}/{dt}^+,M}^2/ \sigma_{{ds}/{dt},M}^{2} \approx 0.9$).

Finally, in the right panel of Fig.\ \ref{fig:converge_ratiocheck}, we plot $P_M(s)$ as computed from eq.\ (\ref{eq:psconditional}), integrating this equation outwards from the center of the distribution using bins of width $0.02$. As shown in \cite{Paper1}, this expression provides a good fit to $P_M(s)$.

\end{document}